\documentstyle[preprint,aps,epsfig]{revtex}
\begin{document}
\draft
\preprint{
\vbox{
\halign{&##\hfil\cr
	& ANL-HEP-PR-99-97 \cr
	& JLAB-THY-99-27 \cr}}
}
\title{Spin Dependence of Massive Lepton Pair Production in Proton-Proton 
Collisions}
\author{Edmond L. Berger$^a$, Lionel E. Gordon$^{b,c}$, and Michael Klasen$^a$}
\address{$^a$High Energy Physics Division,
             Argonne National Laboratory \\
             Argonne, Illinois 60439 \\
	$^b$Jefferson Laboratory, Newport News, VA 23606 \\
	$^c$Hampton University, Hampton, VA 23668}
\date{September 20, 1999}
\maketitle

\begin{abstract} 
We calculate the transverse momentum distribution for the 
production of massive lepton-pairs in longitudinally polarized 
proton-proton reactions at collider energies within the context of 
perturbative quantum chromodynamics.  For values of the transverse 
momentum $Q_T$ greater than roughly half the pair mass $Q$, $Q_T > Q/2$, we 
show that the differential cross section is dominated by subprocesses 
initiated by incident gluons, provided that the polarized gluon density is not 
too small.  Massive lepton-pair differential cross sections should be a good  
source of independent constraints on the polarized gluon density, free from 
the experimental and theoretical complications of photon isolation that beset 
studies of prompt photon production.  We provide predictions for the 
spin-averaged and spin-dependent differential cross sections  
as a function of $Q_T$ at energies relevant for the Relativistic Heavy Ion 
Collider (RHIC) at Brookhaven, and we compare these with predictions for 
real prompt photon production.  
\end{abstract} 
\vspace{0.2in}

\section{Introduction and Motivation}
\label{sec:1}

Both massive lepton-pair production, $h_1 + h_2 \rightarrow \gamma^* + X; 
\gamma^* \rightarrow l \bar{l}$, and prompt real photon production,  
$h_1 + h_2 \rightarrow \gamma + X$ are valuable probes of short-distance 
behavior in hadron reactions.  The two reactions supply critical information 
on parton momentum densities, in addition to the opportunities they offer for 
tests of perturbative quantum chromodynamics (QCD).
Spin-averaged parton momentum densities may be 
extracted from spin-averaged nucleon-nucleon reactions, and spin-dependent 
parton momentum densities from spin-dependent nucleon-nucleon reactions.  An 
ambitious experimental program of measurements of spin-dependence in 
polarized proton-proton reactions will begin soon at Brookhaven's 
Relativistic Heavy Ion Collider (RHIC) with kinematic coverage extending well 
into the regions of phase space in which perturbative quantum chromodynamics 
should yield reliable predictions.  

Massive lepton-pair production, commonly 
referred to as the Drell-Yan process~\cite{ref:DY}, provided early 
confirmation of three colors and of the size of next-to-leading contributions 
to the cross section differential in the pair mass Q.  The mass and 
longitudinal momentum (or rapidity) dependences of the 
cross section (integrated over the transverse momentum $Q_T$ of the pair) 
serve as laboratory for measurement of the {\it antiquark} momentum density, 
complementary to deep-inelastic lepton scattering from which one gains 
information of the sum of the quark and antiquark densities.  Inclusive prompt 
real photon production is a source of essential information on the 
{\it gluon} momentum density.  At lowest order in perturbation theory, the 
reaction is dominated at large values of the transverse momentum $p_T$ of the 
produced photon by the ``Compton" subprocess, $q + g \rightarrow \gamma + q$.  
This dominance is preserved at higher orders, indicating that the experimental 
inclusive cross section differential in $p_T$ may be used to determine the 
density of gluons in the initial 
hadrons~\cite{ref:BQ,ref:Baer,ref:Aurenche,ref:GV}.  

In two previous papers~\cite{ref:BGKDY}, we addressed the production of massive 
lepton-pairs 
as a function of the transverse momentum $Q_T$ of the pair in unpolarized 
nucleon-nucleon reactions, $h_1 + h_2 \rightarrow 
\gamma^* + X$, in the region where $Q_T$ is greater than roughly half of 
the mass of the pair, $Q_T > Q/2$.  We demonstrated that the differential 
cross section in this region is dominated by subprocesses initiated by 
incident gluons.  Correspondingly, massive lepton-pair differential cross 
sections in unpolarized nucleon-nucleon reactions are a valuable, heretofore 
overlooked, independent source of constraints on the spin-averaged gluon 
density. 

Turning to longitudinally polarized proton-proton collisions in this paper, 
we study the potential advantages that the Drell-Yan process may offer 
for the determination of the spin-dependence of the gluon density.  To be 
sure, the cross section for massive lepton-pair production is smaller than 
it is for prompt photon production.  However, just as in the unpolarized case, 
massive lepton pair production is cleaner theoretically since long-range 
fragmentation contributions are absent as are the experimental and 
theoretical complications associated with isolation of the real photon.  
Moreover, the dynamics of spin-dependence in hard-scattering processes is 
a sufficiently complex topic, and its understanding at an early stage in its 
development, that several defensible approaches for extracting polarized 
parton densities deserve to be pursued with the expectation that consistent 
results must emerge.  

There are notable similarities and differences in the theoretical analyses of 
massive lepton-pair production and prompt real photon production.  
At first-order in the strong coupling strength, $\alpha_s$, 
the Compton subprocess and the annihilation subprocess $q +\bar{q} \rightarrow 
\gamma + g$ supply the transverse momentum of the {\it directly} produced 
prompt photons.  Identical subprocesses, with the real $\gamma$ replaced by a
virtual $\gamma^*$, are responsible at ${\cal O}(\alpha_s)$ for the transverse 
momentum of massive lepton-pairs.  An important distinction, however, is that 
fragmentation subprocesses play a very important role in prompt real photon 
production at collider energies.  In these long-distance fragmentation 
subprocesses, the photon emerges from the fragmentation of a final parton, 
e.g., $q + g \rightarrow q + g$, followed by $q \rightarrow \gamma + X$.  
The necessity to invoke phenomenological fragmentation functions and the 
infrared ambiguity~\cite{ref:BGQ} of the isolated cross section in 
next-to-leading 
order raise questions about the extent to which isolated prompt photon 
data may be used for fully quantitative determinations of the gluon density.  
It is desirable to investigate other physical processes for extraction of the 
gluon density that are free from these systematic uncertainties.   
Fortunately, no isolation would seem necessary in the case of virtual photon 
production (and subsequent decay into a pair of muons) in typical collider or 
fixed target experiments.  Muons are observed only after they have penetrated a 
substantial hadron absorber.  Thus, any hadrons within a typical cone about 
the direction of the $\gamma^*$ will have been stopped, and the massive 
lepton-pair signal will be entirely inclusive.

Another significant distinction between 
massive lepton-pair production and prompt real photon production 
is that interest in $h_1 + h_2 \rightarrow 
\gamma^* + X$ has been drawn most often to the domain in which the pair mass 
$Q$ is relatively large, justifying a perturbative treatment based on a small 
value of $\alpha_s(Q)$ and the neglect of inverse-power high-twist 
contributions (except near the edges of phase space).  The focus 
in prompt real photon production is directed to the region of large values 
of $p_T$ where $\alpha_s(p_T)$ is small.  Interest in the transverse 
momentum $Q_T$ dependence of the 
massive lepton-pair production cross section has tended to be limited to 
small values of $Q_T$ where the cross section is largest.  Fixed-order 
perturbation theory~\cite{ref:Reno} is applicable for large $Q_T$, but it is 
inadequate at small $Q_T$, and all-orders resummation 
methods~\cite{ref:CSS,ref:DWS,ref:AEGM,ref:AK,ref:LY} 
have been developed to address the region $Q_T << Q$.  

As long as $Q_T$ is large, the perturbative requirement of small 
$\alpha_s(Q_T)$ can be satisfied without a large value of $Q$.  We therefore 
explore and advocate the potential advantages of studies of 
$d^2\sigma/dQ dQ_T$ as a function of $Q_T$ for modest values of $Q$, 
$Q \sim 2$ to 3 GeV, below the range of the traditional Drell-Yan region.    
There are various backgrounds with which to contend at small $Q$ such as the 
contributions to the event rate from prompt decays of heavy flavors, e.g., 
$h_1 + h_2 \rightarrow c + \bar{c} + X; c \rightarrow l + X$.  These heavy 
flavor contributions may be estimated by direct computation~\cite{ref:BerSop} 
and/or bounded through experimental measurement of the like-sign-lepton 
distributions.
   
In Sec.~II, we review perturbative QCD calculations 
of the transverse momentum distribution for massive lepton-pair production in 
the case in which the initial nucleon spins are polarized as well as in the 
spin-average case.  In Sec.~III, we present next-to-leading order 
predictions for the transverse momentum dependence of the cross sections for 
massive lepton-pair and real prompt photon production in unpolarized 
proton-proton collisions at energies typical of the RHIC collider.   
Predictions for spin dependence are provided in Sec.~IV. 
Our conclusions are summarized in Sec.~V.  


\section{Massive Lepton Pair Production and Prompt Photon Production at 
Next-to-leading Order}
\label{sec:2}

In inclusive hadron interactions at collider energies, $h_1 + h_2 \rightarrow
\gamma^* + X$ with $\gamma^* \rightarrow l \bar{l}$, lepton pair
production proceeds through partonic hard-scattering processes involving
initial-state light quarks $q$ and gluons $g$. In lowest-order QCD, at
${\cal O}(\alpha_s^0)$, the only partonic subprocess is $q + \bar{q}
\rightarrow \gamma^*$. At ${\cal O}(\alpha_s)$, both $q + \bar{q} \rightarrow
\gamma^* + g$ and $q + g \rightarrow \gamma^* + q$ participate, with the
recoil of the final parton balancing the transverse momentum of the
lepton-pair. These processes are shown in Figs.~1(a) and 2(a). Calculations
of the cross section at order ${\cal O}(\alpha_s^2)$ involve virtual loop
corrections to these ${\cal O}(\alpha_s)$ subprocesses (Figs.~1(b) and 2(b))
as well as contributions from a wide range of 
$2 \rightarrow 3$ parton subprocesses (of which some examples are shown in 
Figs.~1(c) and 2(c)).

The physical cross section is obtained through the factorization theorem,
\begin{equation}
 \frac{d^2\sigma_{h_1h_2}^{\gamma^*}}{dQ_T^2dy} = \sum_{ij} \int dx_1 dx_2
 f^i_{h_1}(x_1,\mu_f^2) f^j_{h_2}(x_2,\mu_f^2) 
 \frac{sd^2\hat{\sigma}_{ij}^{\gamma^*}}
 {dtdu}(s,Q,Q_T,y;\mu_f^2).
 \label{dy1}
\end{equation}
It depends on the hadronic center-of-mass energy $S$ and on the mass $Q$, the
transverse momentum $Q_T$, and the rapidity $y$ of the virtual photon; $\mu_f$ 
is the factorization scale of the scattering process.  The
usual Mandelstam invariants in the partonic system are defined by $s =
(p_1+p_2)^2,~t = (p_1-p_{\gamma^*})^2$, and $u = (p_2-p_{\gamma^*})^2$,
where $p_1$ and $p_2$ are the momenta of the initial state partons and
$p_{\gamma^*}$ is the momentum of the virtual photon.  The indices $ij \in
\{q\bar{q},qg\}$ denote the initial parton channels whose contributions are 
added incoherently to yield the total physical cross section.  Functions 
$f^j_{h}(x,\mu)$ denote the usual spin-averaged parton distribution functions.  

The partonic cross
section $\hat\sigma_{ij}^{\gamma^*}(s,Q,Q_T,y;\mu_f^2)$ is obtained commonly
from fixed-order QCD calculations through
\begin{equation}
 \frac{d^2\hat{\sigma}_{ij}^{\gamma^*}}{dtdu} =
   \alpha_s  (\mu^2) \frac{d^2\hat{\sigma}_{ij}^{\gamma^*,(a)}}{dtdu}
 + \alpha_s^2(\mu^2) \frac{d^2\hat{\sigma}_{ij}^{\gamma^*,(b)}}{dtdu}
 + \alpha_s^2(\mu^2) \frac{d^2\hat{\sigma}_{ij}^{\gamma^*,(c)}}{dtdu}
 + {\cal O} (\alpha_s^3).
 \label{dy2}
\end{equation}
The tree, virtual loop, and real emission contributions are labeled
(a), (b), and (c) as are the corresponding diagrams in Figs.~1 and 2.  The 
parameter $\mu$ is the renormalization scale.  It is set equal to the 
factorization scale $\mu_f = \sqrt{Q^2+Q_T^2}$ throughout this paper.

The cross section for $h_1 + h_2 \rightarrow \l\bar{l} + X$, differential
in the invariant mass of the lepton pair $Q^2$ as well as its transverse 
momentum and rapidity, is obtained from Eq.~(\ref{dy1}) by the relation
\begin{equation}
 \frac{d^3\sigma_{h_1h_2}^{l\bar{l}}}{dQ^2dQ_T^2dy} = \left(
 \frac{\alpha_{em}}{3\pi Q^2} \right) \frac{d^2\sigma_{h_1h_2}^{\gamma^*}}
 {dQ_T^2dy}(S,Q,Q_T,y), 
 \label{dy3}
\end{equation}
where $Q^2 = (p_l + p_{\bar l})^2$, and $p_l, p_{\bar l}$ are the four-momenta 
of the two final leptons. 
The Drell-Yan factor $\alpha_{em}/(3\pi Q^2)$ is included in all numerical 
results presented in this paper.

While the full next-to-leading order QCD calculation exists for 
massive lepton-pair production in the case of unpolarized initial nucleons, 
only a partial calculation is available in the polarized case~\cite{ref:CCFG}.  
Correspondingly, we present spin-averaged differential cross sections at 
next-to-leading order, but we calculate spin asymmetries at leading order.  
Spin asymmetries are obtained by dividing the spin-dependent differential 
cross section by its spin-averaged counterpart.  For prompt photon 
production, comparisons of asymmetries computed at next-to-leading order with 
those at leading order show only modest differences~\cite{ref:Frixvogel}, 
whereas the cross sections themselves are affected more significantly.  Given 
the similarity of prompt photon production and massive lepton-pair 
production in the region of $Q_T$ of interest to us~\cite{ref:BGKDY}, we 
expect that the leading-order asymmetries will serve as a useful guide for 
massive lepton-pair production.  

Rewriting Eq.~(\ref{dy3}) and integrating over an interval in $Q^2$, we 
calculate the spin-averaged differential cross section  
$E d^3\sigma^{l\bar{l}}_{h_1 h_2}/dp^3$ as 
\begin{equation}
 \frac{Ed^3\sigma_{h_1 h_2}^{l\bar{l}}}{dp^3} =
\frac{\alpha_{em}}{3\pi^2 S}\sum_{ij}\int^{Q^2_{max}}_{Q^2_{min}}
\frac{dQ^2}{Q^2} \int^1_{x^{min}_1} \frac{dx_1}{x_1-\bar{x}_1} 
 f^i_{h_1}(x_1,\mu_f^2) f^j_{h_2}(x_2,\mu_f^2) 
s \frac{d\hat{\sigma}_{ij}^{\gamma^*}}
 {dt}.
 \label{sdy1}
\end{equation}
In Eq.~(\ref{sdy1}), $Q^2_{max}$ and $Q^2_{min}$ are the chosen upper and 
lower limits of integration for $Q^2$, and 
$x_1^{min}=(\bar{x}_1-\tau)/(1-\bar{x}_2)$.  
The value of $x_2$ is determined from 
$x_2=(x_1\bar{x}_2-\tau)/(x_1-\bar{x}_1)$, with 
\begin{equation}
\bar{x}_1=\frac{Q^2-U}{S}=\frac{1}{2}e^y\sqrt{x_T^2+4\tau} ,
\end{equation}
and
\begin{equation}
\bar{x}_2=\frac{Q^2-T}{S}=\frac{1}{2}e^{-y}\sqrt{x_T^2+4\tau}.
\end{equation}
We use $P_1$ and $P_2$ to denote the 
four-vector momenta of the incident nucleons; $S = (P_1 + P_2)^2$.  The 
invariants in the hadronic system, $T=(P_1-p_{\gamma^*})^2$ and 
$U=(P_2-p_{\gamma^*})^2$, are related to the partonic invariants by 
\begin{equation}
(t-Q^2)=x_1(T-Q^2)=-x_1\bar{x}_2S , 
\end{equation}
and
\begin{equation}
(u-Q^2)=x_2(U-Q^2)=-x_2\bar{x}_1S .
\end{equation}
The scaled variables $x_T$ and $\tau$ are 
\begin{displaymath}
x_T=\frac{2 Q_T}{\sqrt{S}},\;\;\;\; \tau = \frac{Q^2}{S}.  
\end{displaymath}

When the initial nucleons are polarized longitudinally, we can compute 
the difference of cross sections  
\begin{equation}
\Delta \sigma = {\sigma_{++}-\sigma_{+-}} , 
\end{equation}
where $+,-$ denote the helicities of the incident nucleons.  

In analogy to Eq.~(\ref{sdy1}), we find
\begin{equation}
 \frac{Ed^3\Delta \sigma_{h_1 h_2}^{l\bar{l}}}{dp^3} =
\frac{\alpha_{em}}{3\pi^2 S}\sum_{ij}\int^{Q^2_{max}}_{Q^2_{min}}
\frac{dQ^2}{Q^2} \int^1_{x^{min}_1} \frac{dx_1}{x_1-\bar{x}_1} 
\Delta f^i_{h_1}(x_1,\mu_f^2) \Delta f^j_{h_2}(x_2,\mu_f^2) 
s \frac{d\Delta \hat{\sigma}_{ij}^{\gamma^*}}
 {dt}.
 \label{sdy2}
\end{equation}
The functions 
$\Delta f^j_{h}(x,\mu)$ denote the spin-dependent parton distribution 
functions, defined by 
\begin{equation}
\Delta f^i_h(x,\mu_f)=f^i_{h,+}(x,\mu_f)-f^i_{h,-}(x,\mu_f) ;
\end{equation}
$f^i_{h\pm}(x,\mu_f)$ is the distribution of partons of type $i$
with positive $(+)$ or negative $(-)$ helicity in hadron $h$. 
Likewise, the polarized partonic cross section  
$\Delta \hat {\sigma}^{\gamma^*}$ is 
defined by 
\begin{equation}
\Delta \hat{\sigma}^{\gamma^*}=\hat {\sigma}^{\gamma^*}(+,+)-
\hat {\sigma}^{\gamma^*}(+,-), 
\end{equation}
with $+,-$ denoting the helicities of the incoming partons. 

The hard subprocess cross sections in leading order for the unpolarized 
and polarized cases are  
\begin{equation}
s\frac{d\hat{\sigma}_{q\bar{q}}}{dt}=-s\frac{d\Delta\hat{\sigma}_{q\bar{q}}}{dt}
=e_q^2\frac{2\pi\alpha_{em}C_F}{N_C}\frac{\alpha_s}{s}\left[\frac{u}{t}+
\frac{t}{u}+\frac{2Q^2(Q^2-u-t)}{u t}\right],
\end{equation}

\begin{equation}
s\frac{d\sigma_{qg}}{dt}=-e^2_q\frac{\pi
\alpha_{em}}{N_C}\frac{\alpha_s}{s}\left[\frac{s}{t}+\frac{t}{s}+\frac{2Q^2
u}{s t}\right] ,
\end{equation}
and
\begin{equation}
s\frac{d\Delta \sigma_{qg}}{dt}=e^2_q\frac{\pi
\alpha_{em}}{N_C}\frac{\alpha_s}{s}\left[\frac{2u+s}{t}-\frac{2u+t}{s}\right] .
\end{equation}

Our results on the longitudinal spin dependence are expressed in terms of the 
two-spin longitudinal asymmetry $A_{LL}$, defined by 
\begin{equation}
A_{LL} = \frac{\sigma^{\gamma^*}(+,+)-\sigma^{\gamma^*}(+,-)}
{\sigma^{\gamma^*}(+,+)+\sigma^{\gamma^*}(+,-)},
\end{equation}
where $+,-$ denote the helicities of the incoming protons.


\section{Unpolarized Cross Sections}
\label{sec:3}

We turn in this Section to explicit evaluations of the differential 
cross sections as functions of $Q_T$  at collider energies.  
We work in the $\overline{\rm MS}$ renormalization scheme and set the 
renormalization and factorization scales equal.  We employ 
the MRST set of spin-averaged parton densities~\cite{ref:MRST} and a two-loop 
expression for the strong coupling strength $\alpha_s(\mu)$, with five 
flavors and appropriate threshold behavior at $\mu = m_b$; 
$\Lambda^{(4)} = 300$ MeV.  The strong coupling strength 
$\alpha_s$ is evaluated at a hard scale $\mu = \sqrt{Q^2+Q_T^2}$.
We present results for three values of the center-of-mass energy, 
$\sqrt S =$ 50, 200, and 500 GeV.  

For $\sqrt S =$ 200 GeV, we present the invariant inclusive cross section 
$Ed^3\sigma/d p^3$ as a function of $Q_T$ in Fig.~3.  Shown in this figure 
are the $q {\bar q}$ and $q g$ perturbative contributions 
to the cross section at leading order and at next-to-leading order.  
We average the invariant inclusive cross section over the 
rapidity range -1.0 $< y <$ 1.0 and over the mass interval 
5 $<Q<$ 6 GeV.  For
$Q_T<$ 1.5 GeV, the $q {\bar q}$ contribution exceeds that of $q g$ channel. 
However, for values of $Q_T >$ 1.5 GeV, the $q g$ contribution becomes 
increasingly important.  As shown in Fig.~4(a), the $q g$ contribution 
accounts for about 80 \% of the rate once $Q_T \simeq Q$.  The results in 
Fig.~4(a) also demonstrate that subprocesses other than those initiated by the 
$q {\bar q}$ and $q g$ initial channels contribute negligibly.

In Fig.~4(b), we display the fractional contributions to the cross section 
as a function of $Q_T$ for a larger value of Q: 11 $<Q<$ 12 GeV.  In this 
case, the fraction of the rate attributable to $qg$ initiated subprocesses 
again increases with $Q_T$. It becomes 80 \% for $Q_T \simeq Q$.  

For the calculations reported in Figs.~3 and~4(a,b), we chose values of Q in 
the traditional range for studies of massive lepton-pair production, viz., 
above the interval of the $J/\psi$ and $\psi'$ states and either below 
or above the interval of the $\Upsilon's$.  

For Fig.~4(c), we select the interval 2.0 $<Q<$ 3.0 GeV.  In this region, one 
would be inclined to doubt the reliability of leading-twist perturbative 
descriptions of the cross section $d\sigma/dQ$, {\it integrated} over all 
$Q_T$.  However for values of $Q_T$ that are large enough, a perturbative 
description of the $Q_T$ dependence of $d^2\sigma/dQdQ_T$ ought to be 
justified.  The results presented in Fig.~4(c) demonstrate that, as at 
higher masses, the $qg$ incident subprocesses dominate the cross section 
for $Q_T \simeq Q$.

The calculations presented in Figs.~4 show convincingly that data on the 
transverse momentum dependence of the cross section for massive lepton-pair 
production at RHIC collider energies should be a valuable independent source 
of information on the spin-averaged gluon density.  

In Fig.~5, we provide next-to-leading order predictions of the differential 
cross section as a function of $Q_T$ for three values of the center-of-mass 
energy and two intervals of mass $Q$.  Taking  
$Ed^3\sigma/dp^3 = 10^{-3} \rm{pb/GeV}^2$ as the minimum accessible cross 
section, we may use the curves in Fig.~5 to establish that the massive 
lepton-pair cross section may be measured to $Q_T =$ 7.5, 14, and 18.5 GeV at 
$\sqrt S =$ 50, 200, and 500 GeV, respectively, when 2 $< Q <$ 3 GeV, and 
to $Q_T =$ 6, 11.5, and 15 GeV when 5 $< Q <$ 6 GeV.  In terms of reach 
in the fractional momentum $x_{gluon}$ carried by the gluon, these values of 
$Q_T$ may be converted to $x_{gluon} \simeq x_T = 2 Q_T/\sqrt S =$ 0.3, 
0.14, and 0.075 at $\sqrt S =$ 50, 200, and 500 GeV when 2 $< Q <$ 3 GeV, and 
to $x_{gluon} \simeq$ 0.24, 0.115, and 0.06  when 5 $< Q <$ 6 GeV.  On the face 
of it, the smallest value of $\sqrt S$ provides the greatest reach in 
$x_{gluon}$.  However, the reliability of fixed-order perturbative QCD as 
well as dominance of the $qg$ subprocess improve  
with greater $Q_T$.  The maximum value $Q_T \simeq $ 7.5 GeV attainable 
at $\sqrt S = 50$ GeV argues for a larger $\sqrt S$.    
  
It is instructive to compare our results with those 
expected for prompt real photon production.  In Fig.~6, we present the 
predicted differential cross section for prompt photon production for three 
center-of-mass energies.  We display the result with full fragmentation taken 
into consideration (upper line) and with no fragmentation contributions 
included (lower line).  Comparing the magnitudes of the prompt photon and 
massive lepton pair production cross sections in Figs.~5 and~6, we note that 
the inclusive prompt photon cross section is a factor of 1000 to 4000 greater 
than the massive lepton-pair cross section integrated over the mass interval 
2.0 $< Q <$ 3.0 GeV, depending on the value of $Q_T$.  This factor is 
attributable in large measure to the factor $\alpha_{em}/(3 \pi Q^2)$ 
associated with the decay of the virtual photon to $\mu^+ \mu^- $.  
Again taking $Ed^3\sigma/dp^3 = 10^{-3} \rm{pb/GeV}^2$ as the minimum 
accessible cross section, we may use the curves in Fig.~6 to establish that 
the real photon cross section may be measured to $p_T =$ 14, 33, and 52 GeV at 
$\sqrt S =$ 50, 200, and 500 GeV, respectively.  The corresponding reach 
in $x_T = 2 p_T/\sqrt S =$ 0.56, 0.33, and 0.21 at $\sqrt S =$ 50, 200, and 
500 GeV is two to three times that of the massive lepton-pair case.  

The breakdown of the real photon direct cross section at 
$\sqrt S =$ 200 GeV into its $q {\bar q}$ and $qg$ components is presented 
in Fig.~7.  As may be appreciated from a comparison of 
Figs.~4 and~7, dominance of the $qg$ contribution in the massive lepton-pair 
case is as strong as in the prompt photon case.  
The significantly smaller cross section in the case of massive lepton-pair 
production means that the reach in $x_{gluon}$ is restricted to about a factor 
of two to three less, depending on $\sqrt S$ and $Q$, than that potentially 
accessible with prompt photons in the same sample of data.  Nevertheless, it 
is valuable to be able to investigate the gluon density with a process that 
has reduced experimental and theoretical systematic uncertainties from those 
of the prompt photon case.  

In our previous papers~\cite{ref:BGKDY} we compared 
our spin-averaged cross sections with 
available fixed-target and collider data on massive lepton-pair production at 
large values of $Q_T$, and we were able to establish that fixed-order 
perturbative calculations, without resummation, should be reliable for 
$Q_T > Q/2$.  The region of small $Q_T$ and the 
matching region of intermediate $Q_T$ are complicated by some level of 
phenomenological ambiguity.  Within the resummation approach, 
phenomenological non-perturbative functions play a key role in fixing the shape 
of the $Q_T$ spectrum at very small $Q_T$, and matching methods in the 
intermediate region are hardly unique.  For the goals we have in mind, it would 
appear best to restrict attention to the region $Q_T \geq Q/2$.


\section{Predictions for Spin Dependence}
\label{sec:4}

Given theoretical expressions derived in Sec.~II that 
relate the spin-dependent cross section at the hadron level to spin-dependent 
partonic hard-scattering matrix elements and polarized parton densities, we 
must adopt models for spin-dependent parton densities in order to obtain 
illustrative numerical expectations.   For the spin-dependent densities  
that we need, we use the three different parametrizations 
suggested by Gehrmann and Stirling (GS)~\cite{GS}.  We have verified that 
the positivity requirement 
$\left | \Delta f^j_{h}(x,\mu_f)/f^j_{h}(x,\mu_f) \right | \le 1$ is 
satisfied.  

The current deep 
inelastic scattering data do not constrain the polarized gluon density
tightly, and most groups present more than one plausible parametrization.  
Gehrmann and Stirling~\cite{GS} present three such parametrizations, 
labelled GSA, GSB, and GSC.   In the GSA and GSB sets, $\Delta 
G(x,\mu_o)$ is positive for all $x$, whereas in the GSC set 
$\Delta G(x,\mu_o)$ changes sign.  After evolution to $\mu_f^2 = 100$ GeV$^2$, 
$\Delta G(x,\mu_f)$ remains positive for essentially all $x$ in all three sets, 
but its magnitude is small in the GSB and GSC sets.

In this Section, we present two-spin longitudinal asymmetries for 
massive lepton-pair production as a function of transverse momentum.  Results 
are displayed for $pp$ collisions at the center-of-mass
energies $\sqrt{S}=$ 50, 200, and 500 GeV typical of the Brookhaven RHIC 
collider.

In Figs.~8(a-c), we present the two-spin longitudinal asymmetries, 
$A_{LL}$, as a function of  $Q_T$.  As noted earlier, these asymmetries are 
computed in leading-order.  More specifically, we use leading-order partonic 
subprocess cross sections $\hat{\sigma}$ and $\Delta \hat{\sigma}$ with 
next-to-leading order spin-averaged and spin-dependent parton densities and a 
two-loop expression for $\alpha_s$.  The choice of a leading-order 
expression for $\Delta \hat{\sigma}$ is required because the 
full next-to-leading order derivation of $\Delta \hat{\sigma}$ has not been 
completed for massive lepton-pair production.  Experience with prompt photon 
production indicates that the leading-order and next-to-leading order results 
for the asymmetry are similar so long as both are dominated by the $qg$ 
subprocess.   Results are shown for three 
choices of the polarized gluon density.  The asymmetry becomes sizable for 
large enough $Q_T$ for the GSA and GSB parton sets but not in the GSC case.  
Comparing the three figures, we note that $A_{LL}$ is nearly independent 
of the pair mass $Q$ as long as $Q_T$ is not too small.  This feature should 
be helpful for the accumulation of statistics; small bin-widths in mass 
are not necessary, but the $J/\psi$ and $\Upsilon$ resonance regions should 
be excluded.    

As noted above the $qg$ subprocess dominates the {\it{spin-averaged}} cross 
section. It is interesting and important to inquire whether this dominance 
persists in the spin-dependent situation.  In Figs.~9 and 10, we compare 
the contribution to the asymmetry from the polarized $qg$ subprocess 
with the complete answer for all three sets of parton densities.  The $qg$ 
contribution is more positive than the full answer for values of $Q_T$ that 
are not too small; the 
full answer is reduced by the negative contribution from the 
$q \bar{q}$ subprocess for which the parton-level asymmetry 
${\hat{a}}_{LL} = -1$.  At small $Q_T$, the net 
asymmetry may be driven negative by the $q \bar{q}$ contribution, and based on 
our experience with other calculations~\cite{ref:BergerGordon}, from 
processes such as $gg$ that 
contribute in next-to-leading order.  For the GSA and GSB sets, we see that 
once it becomes sizable (e.g., 5\% or more), the total asymmetry from all 
subprocesses is dominated by the large contribution from the $qg$ subprocess. 

As a general rule in studies of polarization phenomena, many subprocesses can 
contribute small and conflicting asymmetries.  Asymmetries are readily 
interpretable only in situations where the basic dynamics is dominated by one 
major subprocess and the overall asymmetry is sufficiently large.  In the case 
of massive lepton-pair production that is the topic of this paper, when the 
overall asymmetry $A_{LL}$ itself is small, the contribution from the 
$qg$ subprocess cannot be said to dominate the answer.  However, if 
a large asymmetry is measured, similar to that expected in the GSA case at the 
larger values of $Q_T$, Figs.~9 and 10 show that the answer is dominated by 
the $qg$ contribution, and data will serve to 
constrain $\Delta G(x,\mu_f)$.  If $\Delta G(x,\mu_f)$ is small 
and a small asymmetry is measured, such as for the GSC parton set, or at small 
$Q_T$ for all parton sets, one will not be 
able to conclude which of the subprocesses is principally responsible, and no 
information could be adduced about $\Delta G(x,\mu_f)$, except that it 
is small.  

In Figs.~10 (a) and (b), we examine the energy dependence of our predictions 
for two different intervals of mass $Q$.  For $Q_T$ not too small, we observe 
that $A_{LL}$ in massive lepton pair production is well described by a 
scaling function of $x_T = 2Q_T/\sqrt S$, 
$A_{LL}(\sqrt S,Q_T) \simeq h_{\gamma^*}(x_T)$.  
In our discussion of the spin-averaged cross sections, we took
$Ed^3\sigma/dp^3 = 10^{-3} \rm{pb/GeV}^2$ as the minimum accessible cross.  
Combining the results in Fig.~5 with those in Fig.~10, we see that longitudinal 
asymmetries $A_{LL} = 20 \%,\, 7.5 \%,\; \rm{and}\, 3 \%$ are predicted at this 
level of cross section at $\sqrt{S}=$ 50, 200, and 500 GeV when 2 $< Q <$ 3 
GeV, and 
$A_{LL} = 11 \%,\, 5 \%,\, \rm{and}\; 2 \%$ when 5 $< Q <$ 6 GeV.  For a given 
value of $Q_T$, smaller values of $\sqrt S$ result in greater asymmetries 
because $\Delta G(x)/G(x)$ grows with $x$.  

The predicted cross sections 
in Fig.~5 and the predicted asymmetries in Fig.~10 should make it possible 
to optimize the choice of center-of-mass energy at which measurements might 
be carried out.  At $\sqrt{S}=$ 500 GeV, asymmetries are not appreciable in 
the interval of $Q_T$ in which event rates are appreciable.  At the other 
extreme, the choice of $\sqrt{S}=$ 50 GeV does not allow a sufficient range in 
$Q_T$.  Accelerator physics considerations favor higher energies since the 
instantaneous luminosity increases with $\sqrt{S}$.  Investigations in 
the energy interval $\sqrt{S}=$ 150 to 200 GeV would seem preferred.  

In Fig.~11, we display predictions for $A_{LL}$ in prompt real photon 
production for three values of the center-of-mass energy.  These 
calculations are done at next-to-leading order in QCD.  Dominance of the 
$qg$ contribution is again evident as long as $A_{LL}$ is not too small.  So 
long as $Q_T \ge Q$, we note that the asymmetry in massive lepton-pair 
production is about the same size as that in prompt real photon production, 
as might be expected from the strong similarity of the production dynamics in 
the two cases.  As in massive lepton-pair production, $A_{LL}$ in prompt 
photon production is well described by a scaling function of 
$x_T = 2p_T/\sqrt S$, $A_{LL}(\sqrt S, p_T) \simeq h_{\gamma}(x_T)$.  
For $Ed^3\sigma/dp^3 = 10^{-3} \rm{pb/GeV}^2$, we predict 
longitudinal asymmetries $A_{LL} = 31 \%,\, 17 \%,\, \rm{and}\; 10 \%$ in 
real prompt photon production at $\sqrt S =$ 50, 200, and 500 GeV.  


\section{Discussion and Conclusions}
\label{sec:5}

In this paper we focus on the $Q_T$ distribution for $p + p \rightarrow 
\gamma^* + X$.  We present and discuss calculations carried out in 
QCD at RHIC collider energies.  We show that the differential cross section in 
the region $Q_T \geq Q/2$ is dominated by subprocesses initiated by incident 
gluons.  Dominance of the $qg$ contribution in the massive lepton-pair case 
is as strong as in the prompt photon case, $p + p \rightarrow \gamma + X$.
As our calculations demonstrate, the $Q_T$ distribution of massive lepton 
pair production offers a valuable additional method for direct measurement of 
the gluon momentum distribution.  The method is similar in 
principle to the approach based on prompt photon production, but it avoids 
the experimental and theoretical complications of photon isolation that beset 
studies of prompt photon production.  

As long $Q_T$ is large, the perturbative requirement of small $\alpha_s(Q_T)$ 
can be satisfied without a large value of $Q$.  We therefore explore and 
advocate the potential advantages of studies of $d^2\sigma/dQ dQ_T$ as a 
function of $Q_T$ for modest values of $Q$,  
$Q \sim 2$~GeV, below the traditional Drell-Yan region. 
   
For the goals we have in mind, it would 
appear best to restrict attention to the region in $Q_T$ above the value at 
which the resummed result falls below the fixed-order perturbative expectation. 
A rough rule-of-thumb based on our calculations is $Q_T \geq Q/2$.  
Uncertainties associated with resummation make it impossible to use data 
on the $Q_T$ distribution at small $Q_T$ to extract precise information 
on parton densities.  

In this paper we also present a calculation of the 
longitudinal spin-dependence of massive lepton-pair production at 
large values of transverse momentum.  We provide polarization 
asymmetries as functions of transverse momenta that may be 
useful for estimating the feasibility of measurements of spin-dependent cross 
sections in future experiments at RHIC collider energies. 
The Compton subprocess dominates the dynamics in longitudinally 
polarized proton-proton reactions as long as the polarized gluon density 
$\Delta G(x,\mu_f)$ is not too small.  As a result, two-spin measurements of 
inclusive prompt photon production in polarized $pp$ scattering should 
constrain the size, {\it {sign}}, and Bjorken $x$ dependence of 
$\Delta G(x,\mu_f)$.  Significant values of $A_{LL}$ (i.e., greater than 5 \%) 
may be expected for $x_T = 2Q_T/ \sqrt S > 0.10$ if the polarized gluon density 
$\Delta G(x,\mu_f)$ is as large as that in the GSA set of polarized parton 
densities.  If so, the data could be used to determine the polarization of the 
gluon density in the nucleon.  On the other hand, for small $\Delta G(x,\mu_f)$, 
dominance of the $qg$ subprocess is lost, and $\Delta G(x,\mu_f)$ is 
inaccessible.  


\section*{Acknowledgments}

Work in the High 
Energy Physics Division at Argonne National Laboratory is supported by 
the U.S. Department of Energy, Division of High Energy Physics, 
Contract W-31-109-ENG-38.  This work was supported in part by DOE contract 
DE-AC05-84ER40150 under which
Southeastern Universities Research Association operates the Thomas
Jefferson National Accelerator Facility.



\begin{figure}
 \begin{center}
  {\unitlength1cm
  \begin{picture}(15,17)
   \epsfig{file=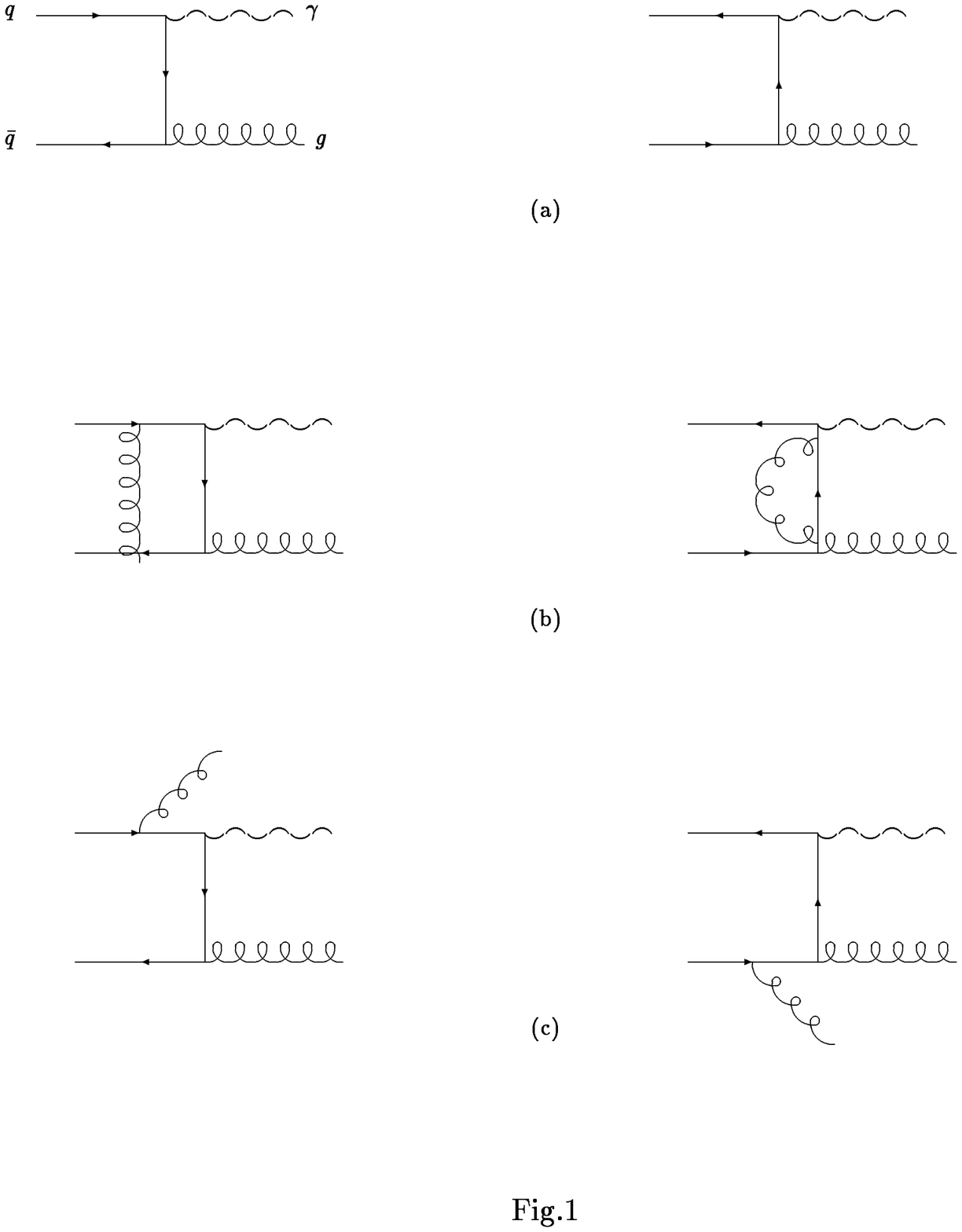,bbllx=95pt,bblly=155pt,bburx=540pt,bbury=645pt,%
           height=17cm,clip=}
  \end{picture}}
 \end{center}
\caption{(a)Lowest-order Feynman diagrams for the direct process 
$q + \bar{q} \rightarrow \gamma + g$. (b) Examples of virtual gluon loop 
diagrams.  (c) Examples of next-to-leading order three-body final-state 
diagrams.}
\label{fig1}
\end{figure}

\begin{figure}
 \begin{center}
  {\unitlength1cm
  \begin{picture}(15,20)
   \epsfig{file=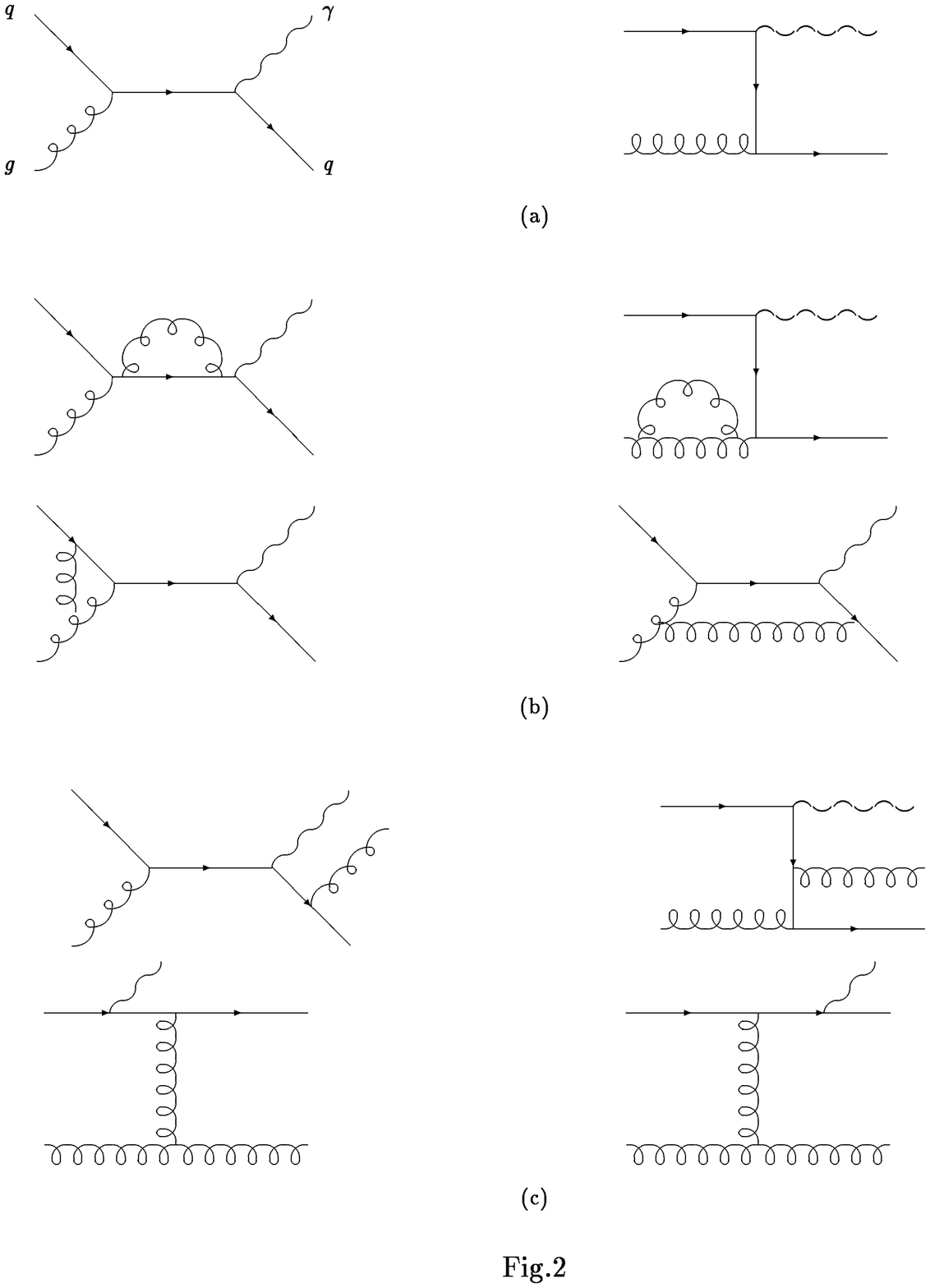,bbllx=85pt,bblly=60pt,bburx=540pt,bbury=655pt,%
           height=20cm,clip=}
  \end{picture}}
 \end{center}
\caption{As in Fig.~1, but for the subprocesses initiated by the $q + g$ 
initial state.}
\label{fig2}
\end{figure}

\begin{figure}
 \begin{center}
  {\unitlength1cm
  \begin{picture}(12,18)
   \epsfig{file=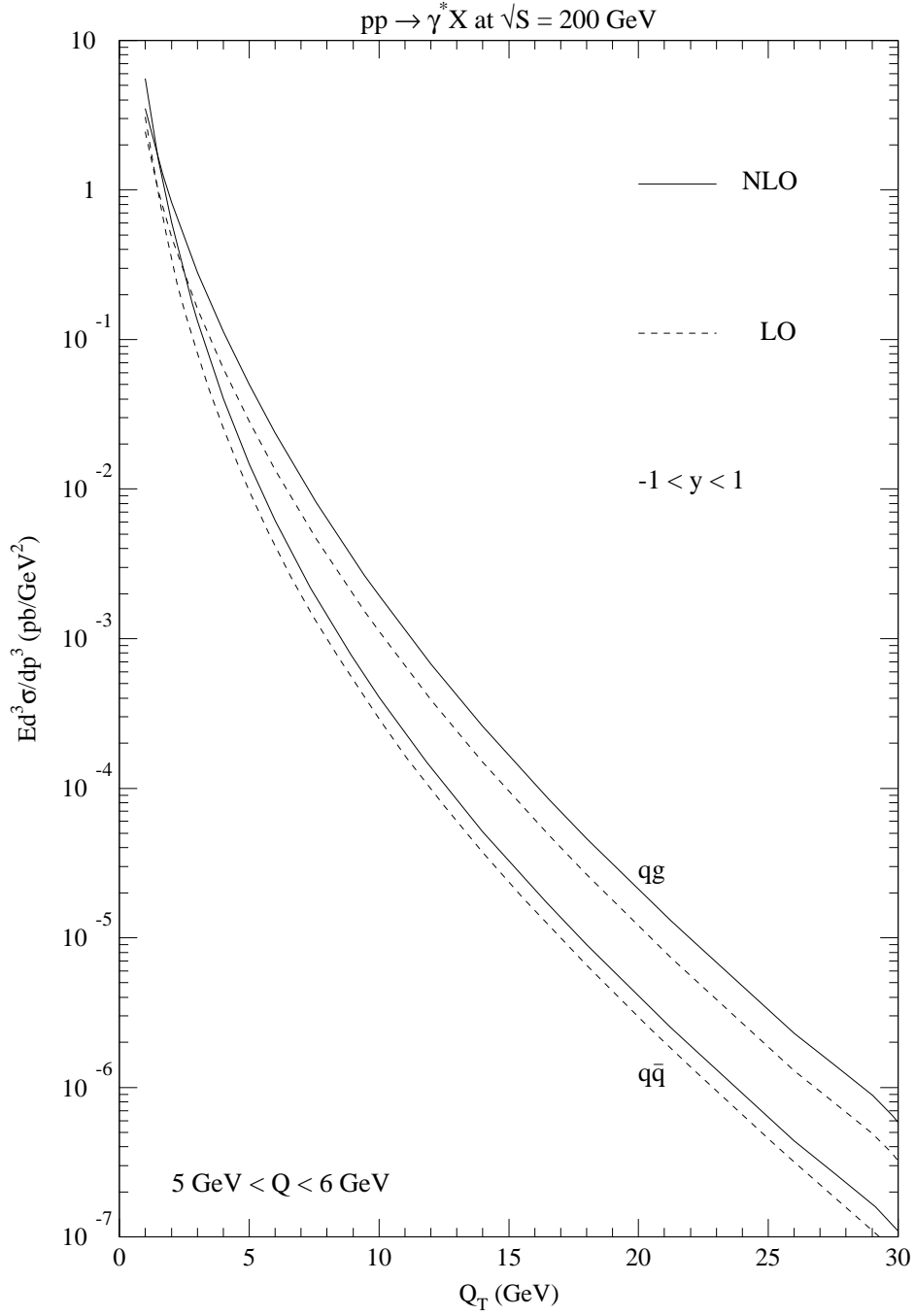,bbllx=55pt,bblly=100pt,bburx=495pt,bbury=725pt,%
           height=18cm}
  \end{picture}}
 \end{center}
\caption{Lowest order (dashed lines) and next-to-leading 
order (solid lines) perturbative calculations of the 
invariant inclusive cross section $Ed^3\sigma/dp^3$ as a function of $Q_T$ for 
$p p \rightarrow \gamma^* X$ at $\sqrt S =$  200 GeV, in the 
$\overline{\rm MS}$ scheme.  Contributions from the $qg$ and $q \bar{q}$ 
channels are shown separately.  The results are averaged over the rapidity 
interval - 1.0 $< y <$ 1.0 and over the interval 
5.0 $<Q<$ 6.0 GeV.}
\label{fig3}
\end{figure}

\begin{figure}
 \begin{center}
  {\unitlength1cm
  \begin{picture}(12,18)
   \epsfig{file=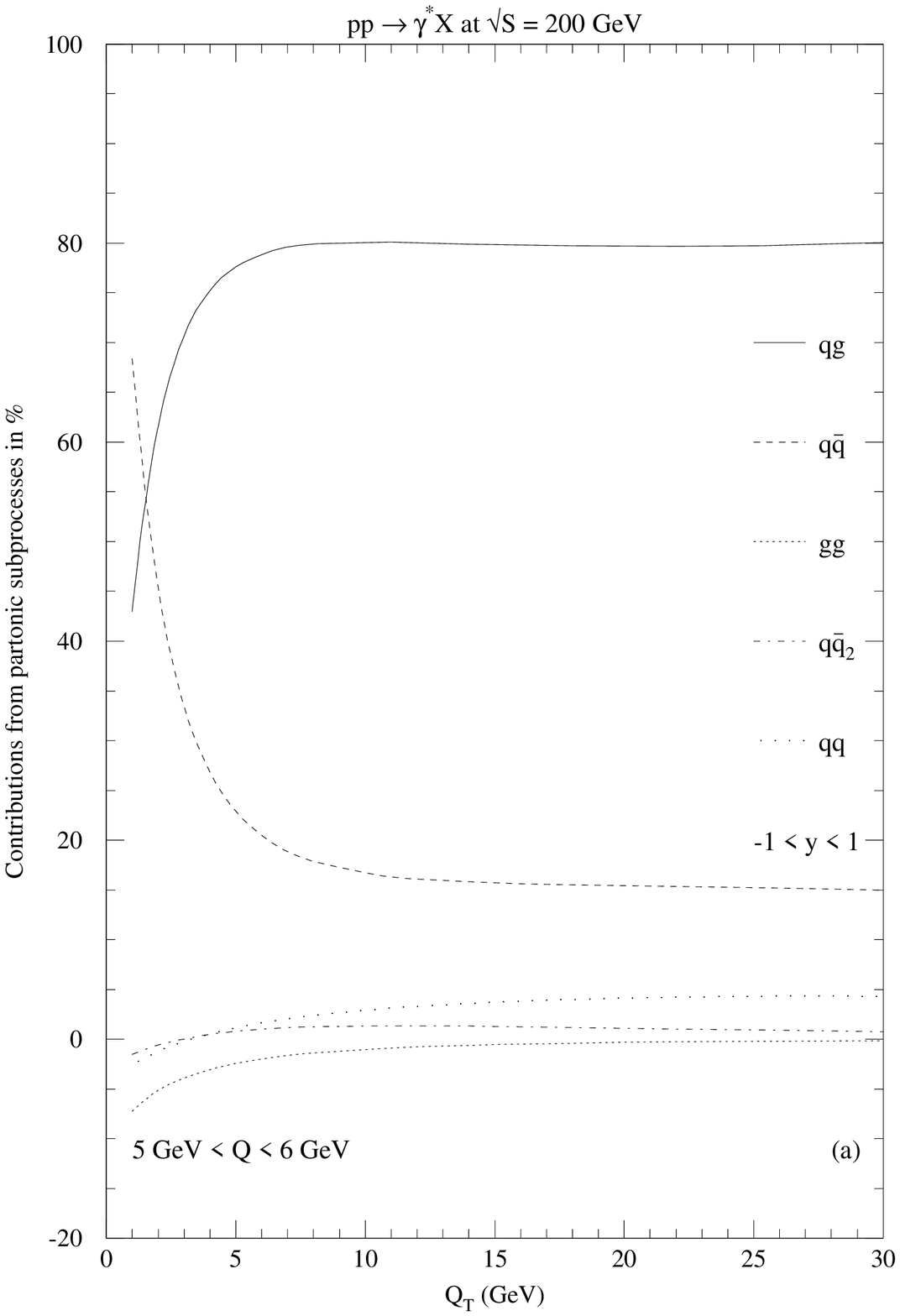,bbllx=55pt,bblly=100pt,bburx=495pt,bbury=725pt,%
           height=18cm}
  \end{picture}}
 \end{center}
 \begin{center}
  {\unitlength1cm
  \begin{picture}(12,18)
   \epsfig{file=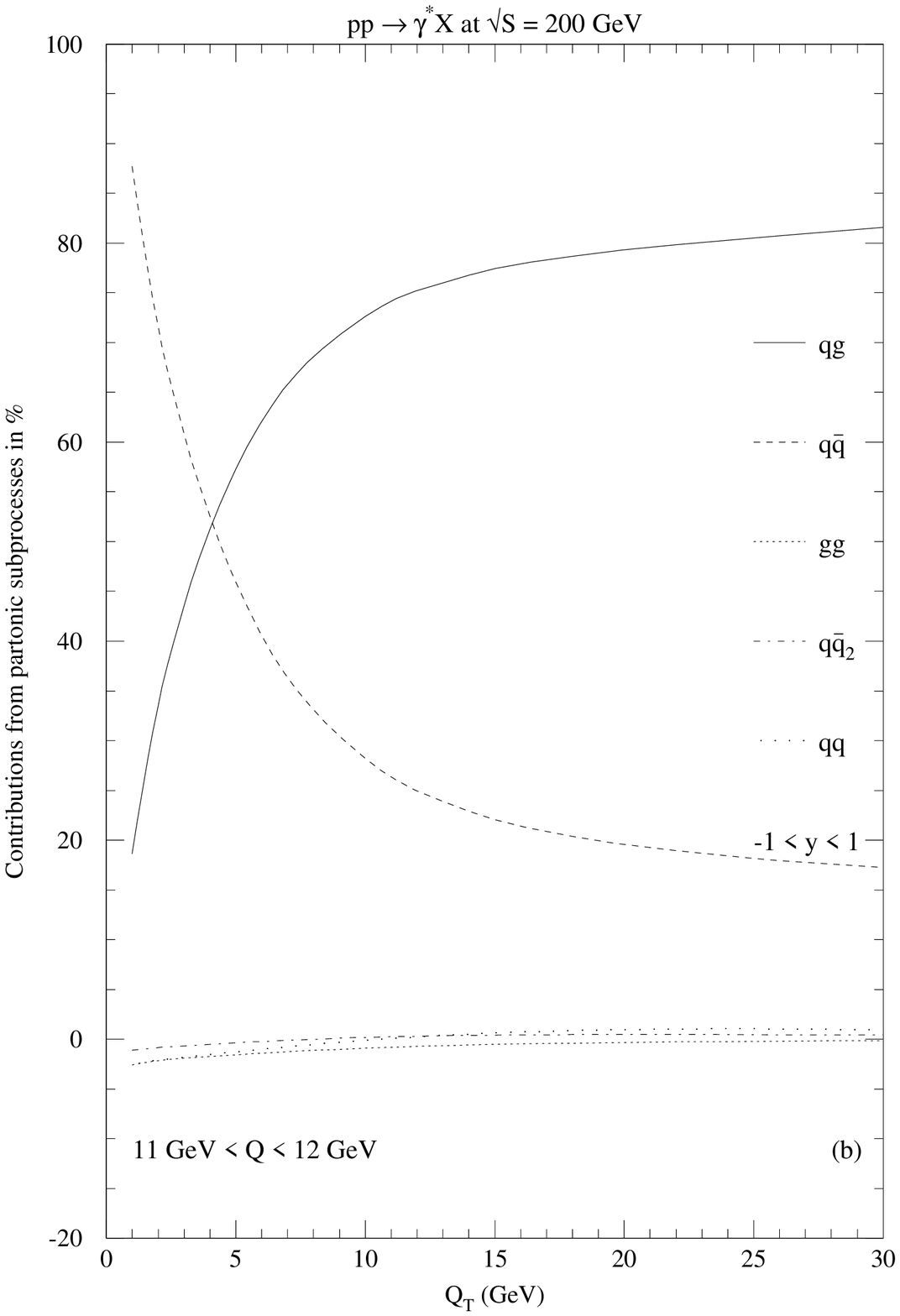,bbllx=55pt,bblly=100pt,bburx=495pt,bbury=725pt,%
           height=18cm}
  \end{picture}}
 \end{center}
 \begin{center}
  {\unitlength1cm
  \begin{picture}(12,18)
   \epsfig{file=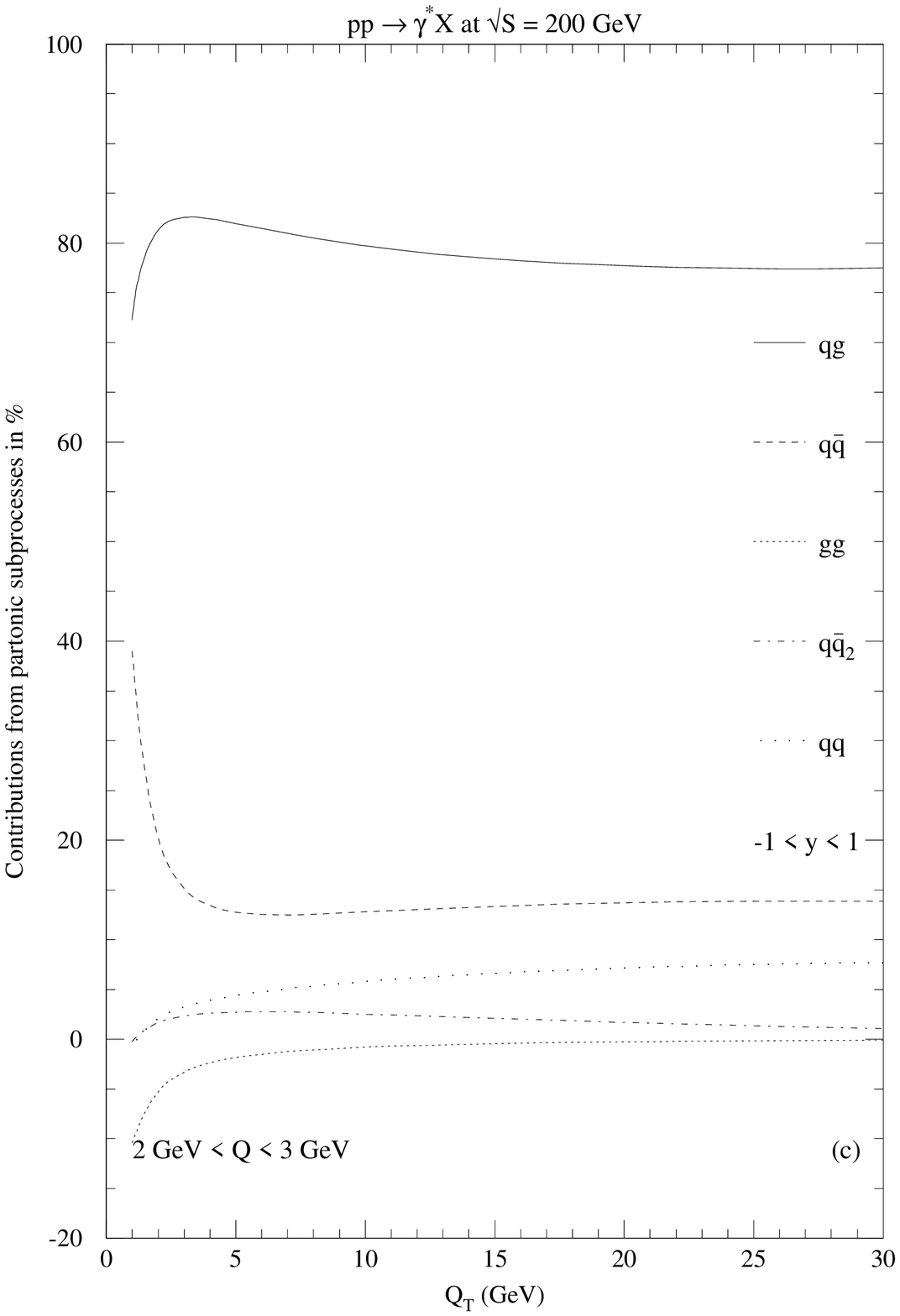,bbllx=55pt,bblly=100pt,bburx=495pt,bbury=725pt,%
           height=18cm}
  \end{picture}}
 \end{center}
\caption{Contributions from the various partonic subprocesses to the invariant 
inclusive cross section $Ed^3\sigma/dp^3$ as a function of $Q_T$ for 
$p p \rightarrow \gamma^* X$ at $\sqrt S =$ 200 GeV.  The cross section is 
averaged over the rapidity interval -1.0 $< y <$ 1.0 and over the intervals 
(a) 5.0 $<Q<$ 6.0 GeV, (b) 11.0 $<Q<$ 12.0 GeV, and (c) 2.0 $<Q<$ 3.0 GeV.
The contributions are labeled by $qg$ (solid), $q \bar{q}$ (dashed), 
$gg$ (dotted), $q \bar{q}_2$ non-factorizable parts (dot-dashed), and $qq$ 
(wide dots).}
\label{fig4}
\end{figure}

\begin{figure}
 \begin{center}
  {\unitlength1cm
  \begin{picture}(12,18)
   \epsfig{file=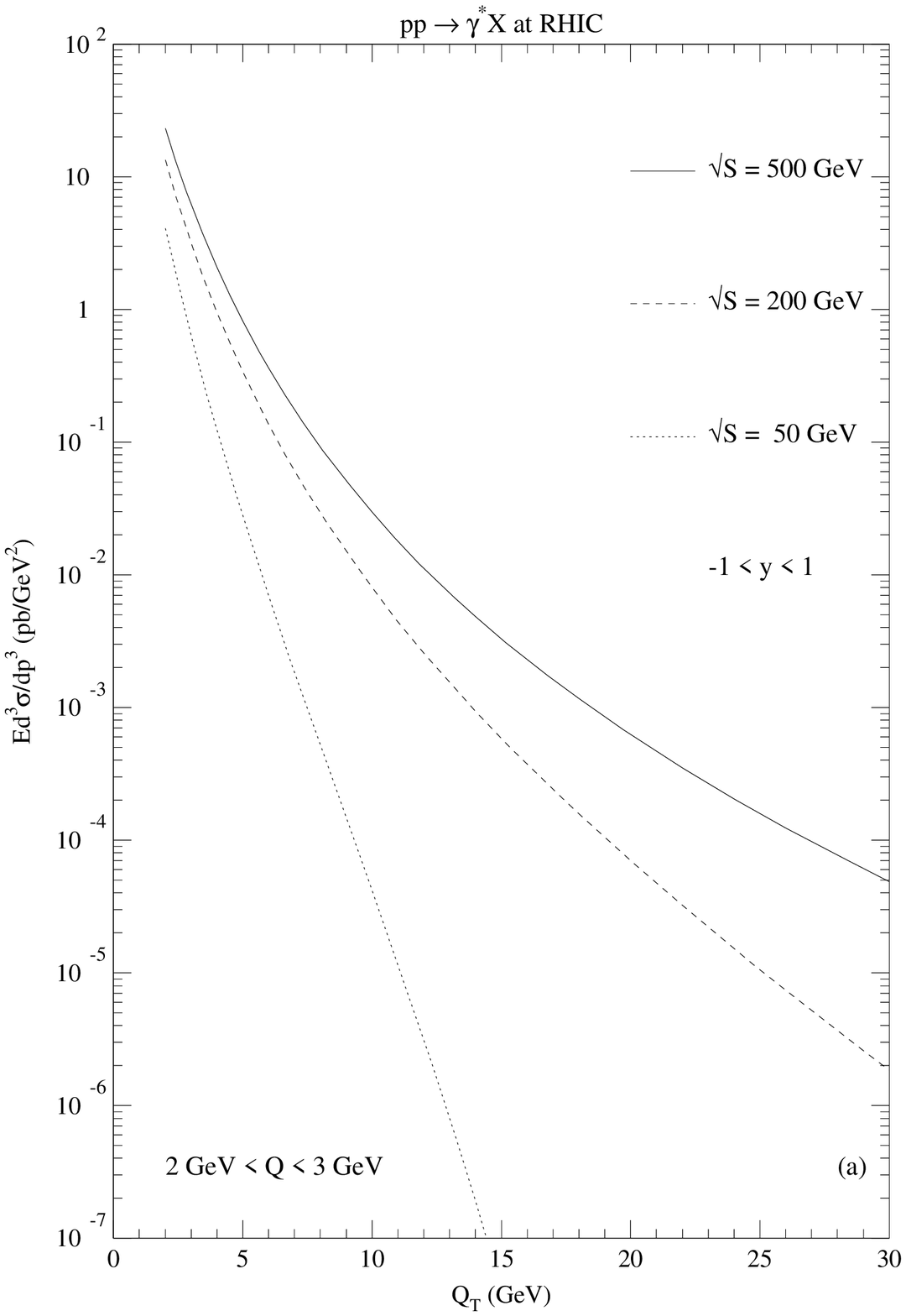,bbllx=55pt,bblly=100pt,bburx=495pt,bbury=725pt,%
           height=18cm}
  \end{picture}}
 \end{center}
 \begin{center}
  {\unitlength1cm
  \begin{picture}(12,18)
   \epsfig{file=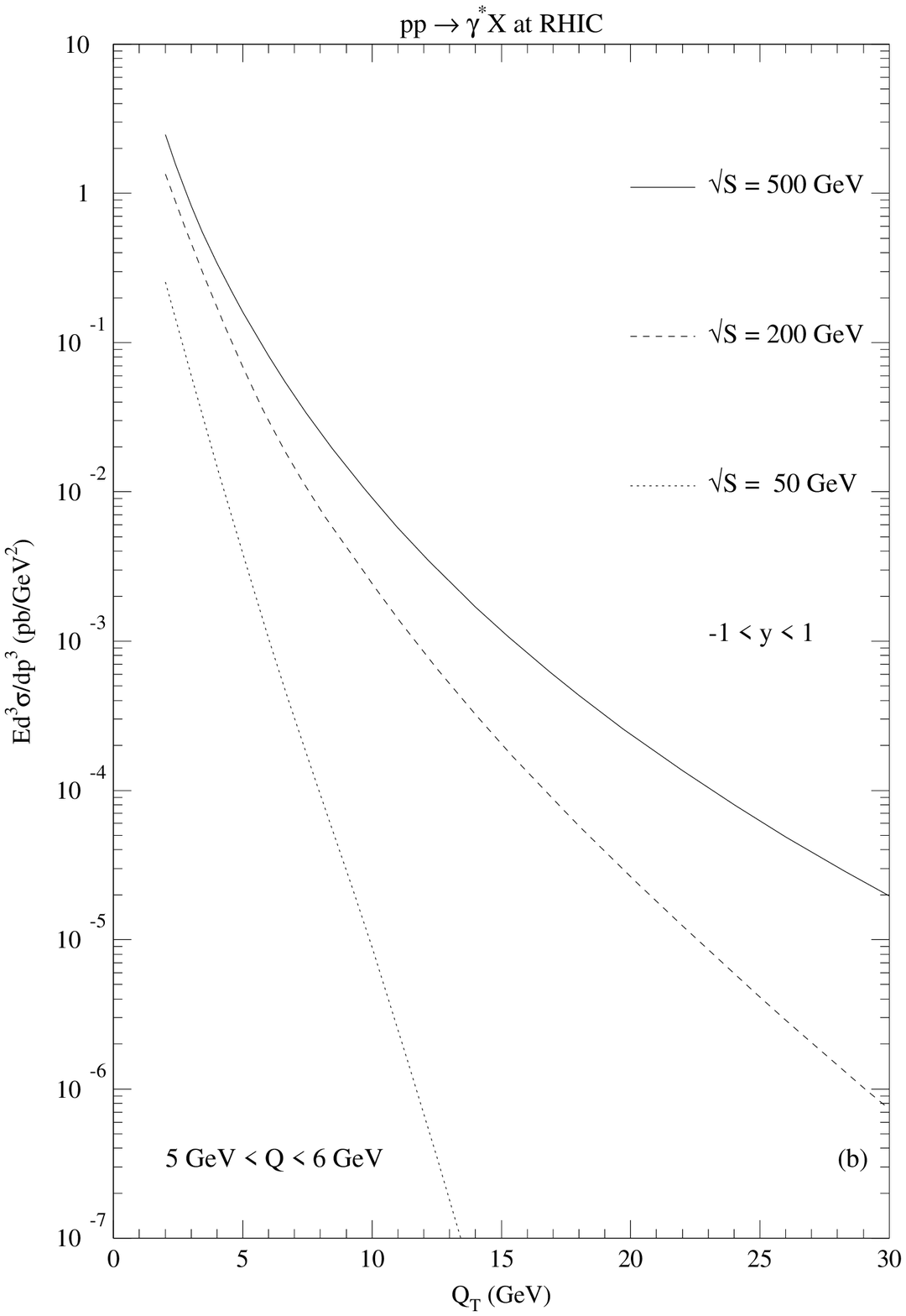,bbllx=55pt,bblly=100pt,bburx=495pt,bbury=725pt,%
           height=18cm}
  \end{picture}}
 \end{center}
\caption{Differential cross sections $Ed^3\sigma/dp^3$ as a function of $Q_T$ 
for for $p p \rightarrow \gamma^* +X$ at $\sqrt S =$  
50, 200, and 500 GeV, averaged over the rapidity interval -1.0 $< y <$ 1.0 
and the mass intervals (a) 2.0 $<Q<$ 3.0 GeV, and (b) 5.0 $<Q<$ 6.0 GeV.}  
\label{fig5}
\end{figure}

\begin{figure}
 \begin{center}
  {\unitlength1cm
  \begin{picture}(12,18)
   \epsfig{file=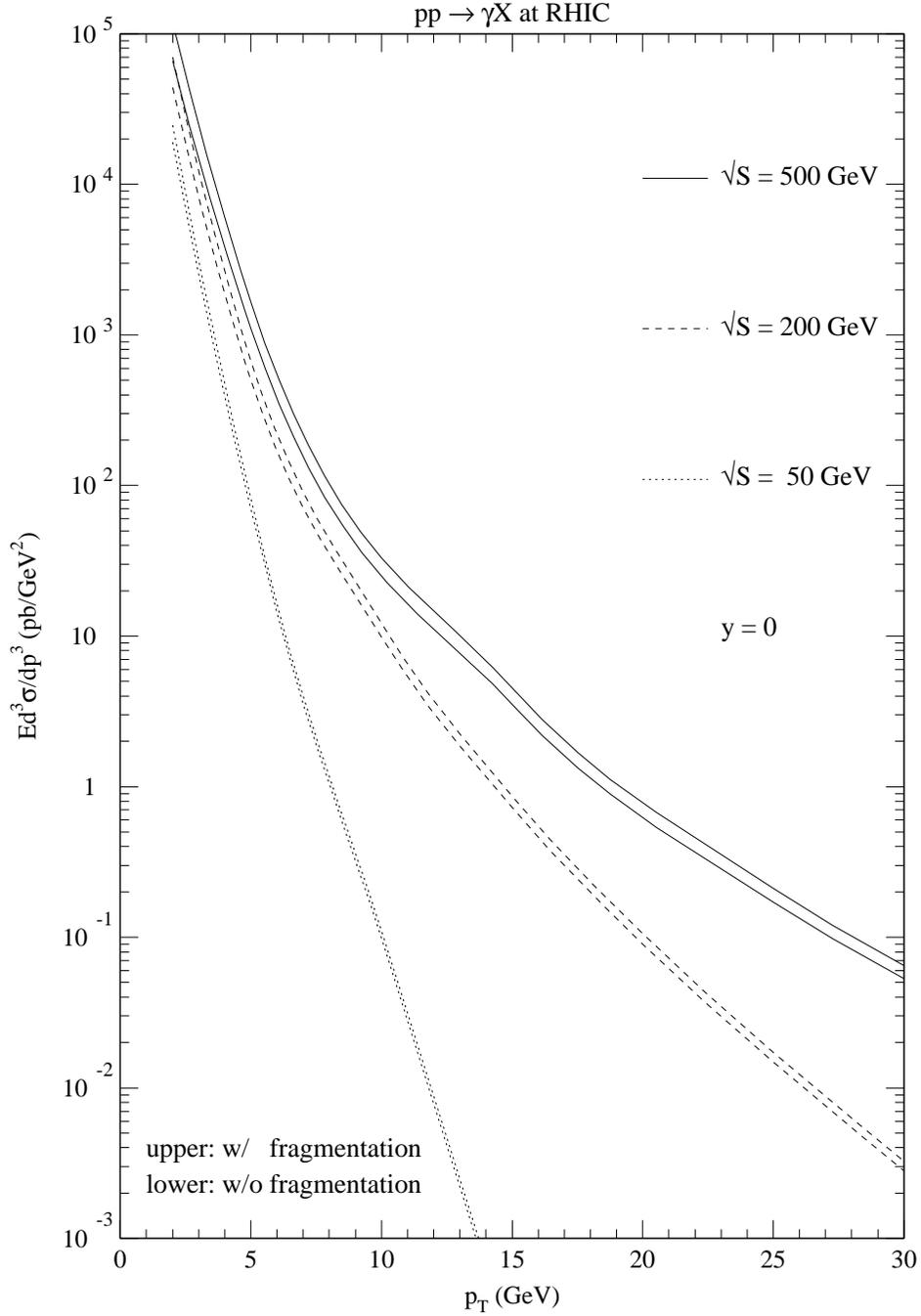,bbllx=55pt,bblly=100pt,bburx=495pt,bbury=725pt,%
           height=18cm}
  \end{picture}}
 \end{center}
\caption{Differential cross sections $Ed^3\sigma/dp^3$ at rapidity $y =$ 0 as 
a function of $p_T$ for real photon production $p p \rightarrow \gamma X$ at 
$\sqrt S = $ 50, 200, and 500 GeV for two 
cases: no fragmentation terms included (lower) and the inclusive case 
with full fragmentation included (upper).}
\label{fig6}
\end{figure}

\begin{figure}
 \begin{center}
  {\unitlength1cm
  \begin{picture}(12,18)
   \epsfig{file=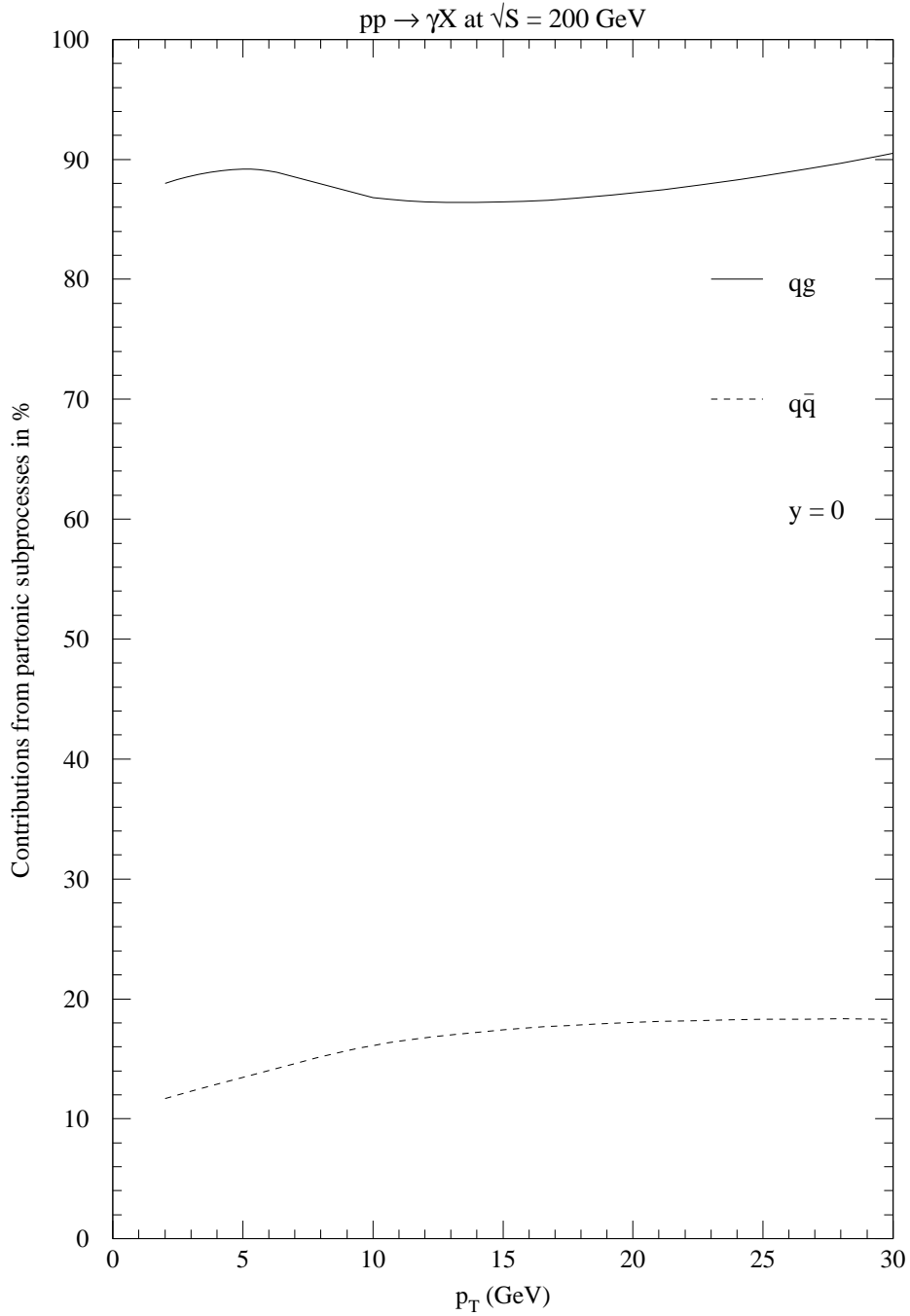,bbllx=55pt,bblly=100pt,bburx=495pt,bbury=725pt,%
           height=18cm}
  \end{picture}}
 \end{center}
\caption{Fractional contributions from the next-to-leading order 
$q\bar{q}$ and the $qg$ channels
to the differential cross sections $Ed^3\sigma/dp^3$ as a function of $p_T$  
for real photon production $p p \rightarrow \gamma X$ at 
$\sqrt S=200$ GeV and rapidity $y =$ 0.  No fragmentation terms are included.}
\label{fig7}
\end{figure}

\begin{figure}
 \begin{center}
  {\unitlength1cm
  \begin{picture}(12,18)
   \epsfig{file=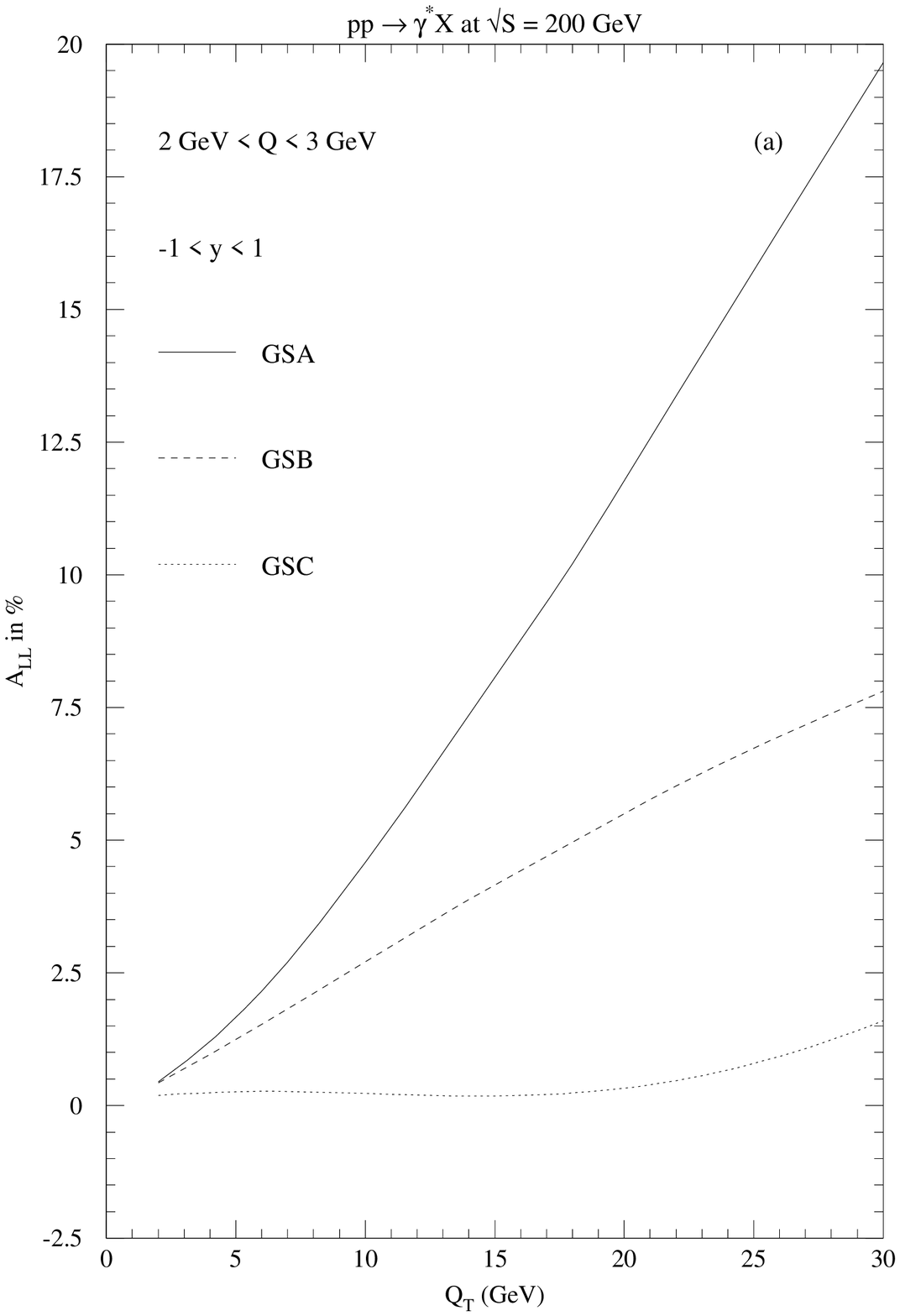,bbllx=55pt,bblly=100pt,bburx=495pt,bbury=725pt,%
           height=18cm}
  \end{picture}}
 \end{center}
 \begin{center}
  {\unitlength1cm
  \begin{picture}(12,18)
   \epsfig{file=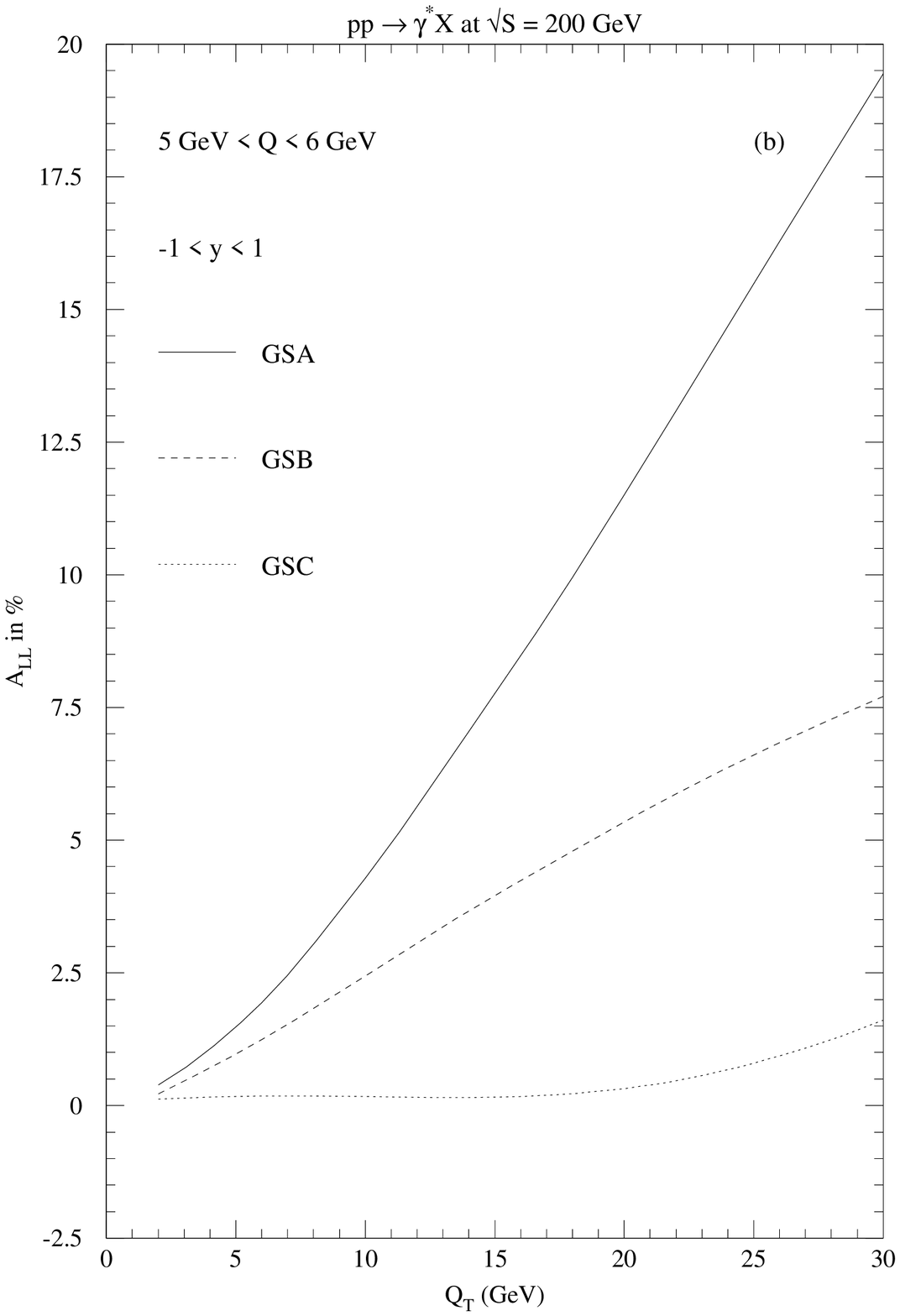,bbllx=55pt,bblly=100pt,bburx=495pt,bbury=725pt,%
           height=18cm}
  \end{picture}}
 \end{center}
 \begin{center}
  {\unitlength1cm
  \begin{picture}(12,18)
   \epsfig{file=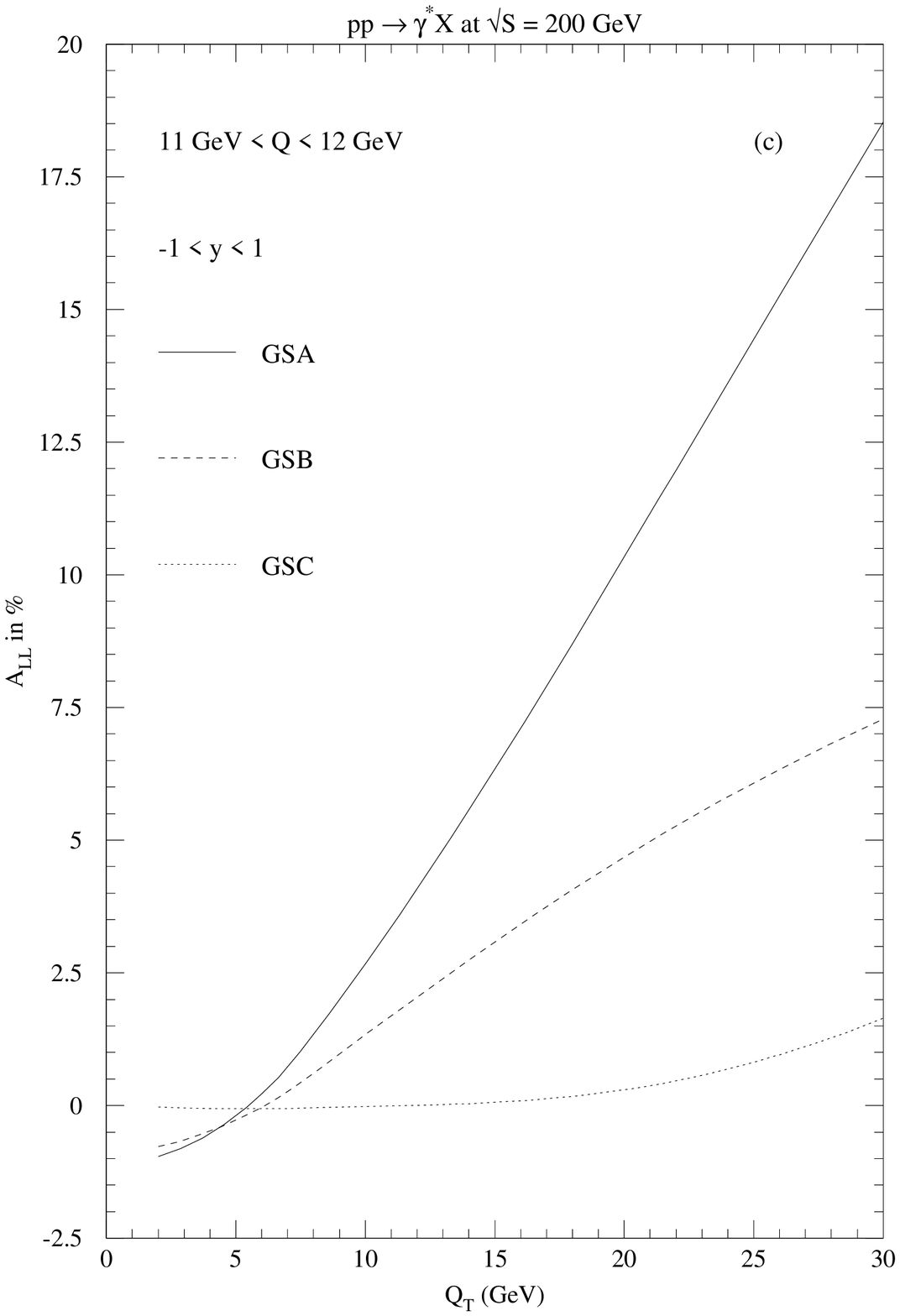,bbllx=55pt,bblly=100pt,bburx=495pt,bbury=725pt,%
           height=18cm}
  \end{picture}}
 \end{center}
\caption{Computed longitudinal asymmetry $\protect A_{LL}$ as a function 
of $Q_T$ for 
$p p \rightarrow \gamma^* X$ at $\sqrt S =$ 200 GeV.  The asymmetry is 
averaged over the rapidity interval -1.0 $< y <$ 1.0 and over the intervals 
(a) 2.0 $<Q<$ 3.0 GeV, (b) 5.0 $<Q<$ 6.0 GeV, and (c) 11.0 $<Q<$ 12.0 GeV, 
for three choices of the polarized parton densities, GSA, GSB, and GSC.}  
\label{fig8}
\end{figure}

\begin{figure}
 \begin{center}
  {\unitlength1cm
  \begin{picture}(12,18)
   \epsfig{file=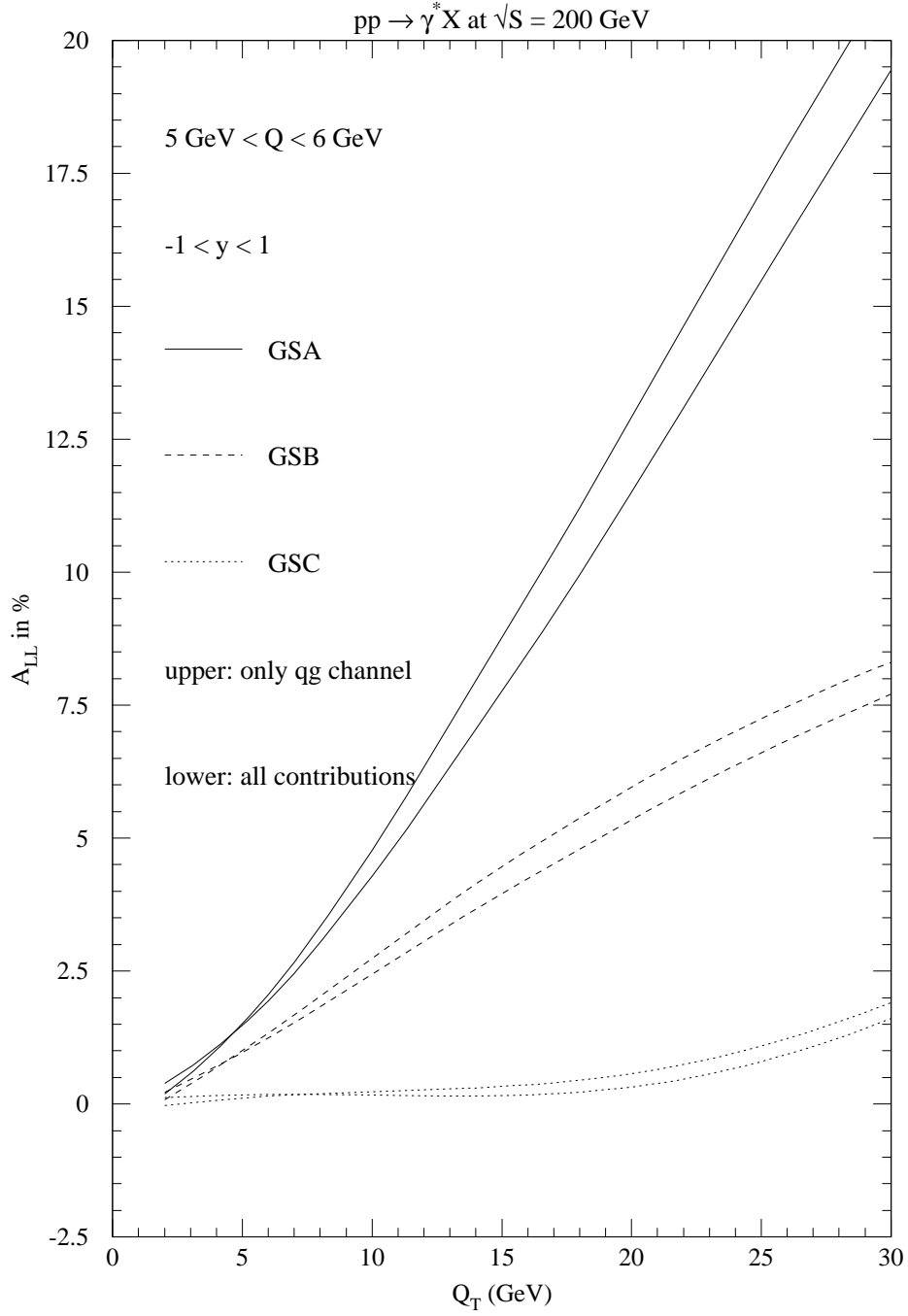,bbllx=55pt,bblly=100pt,bburx=495pt,bbury=725pt,%
           height=18cm}
  \end{picture}}
 \end{center}
\caption{Comparison of the contribution of the $qg$ subprocess (upper curve) 
to the longitudinal asymmetry $\protect A_{LL}$ with the total (lower curve) 
as a function of $Q_T$ for
$p p \rightarrow \gamma^* X$ at $\sqrt S =$ 200 GeV.  The asymmetry is 
averaged over the rapidity interval -1.0 $< y <$ 1.0 and over the interval 
5.0 $<Q<$ 6.0 GeV.  Results are shown for three sets of spin-dependent parton 
densities.}
\label{fig9}
\end{figure}

\begin{figure}
 \begin{center}
  {\unitlength1cm
  \begin{picture}(12,18)
   \epsfig{file=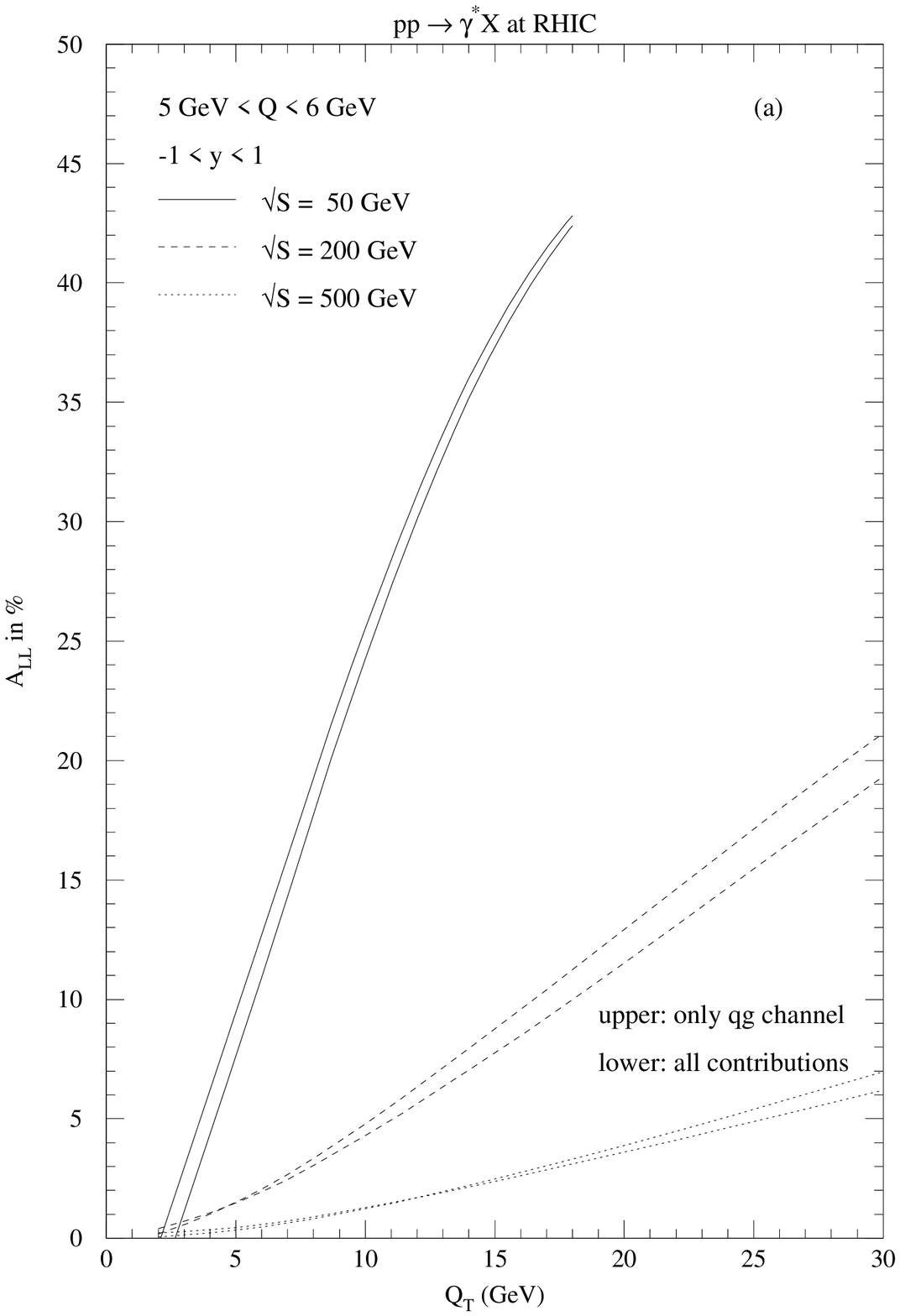,bbllx=55pt,bblly=100pt,bburx=495pt,bbury=725pt,%
           height=18cm}
  \end{picture}}
 \end{center}
 \begin{center}
  {\unitlength1cm
  \begin{picture}(12,18)
   \epsfig{file=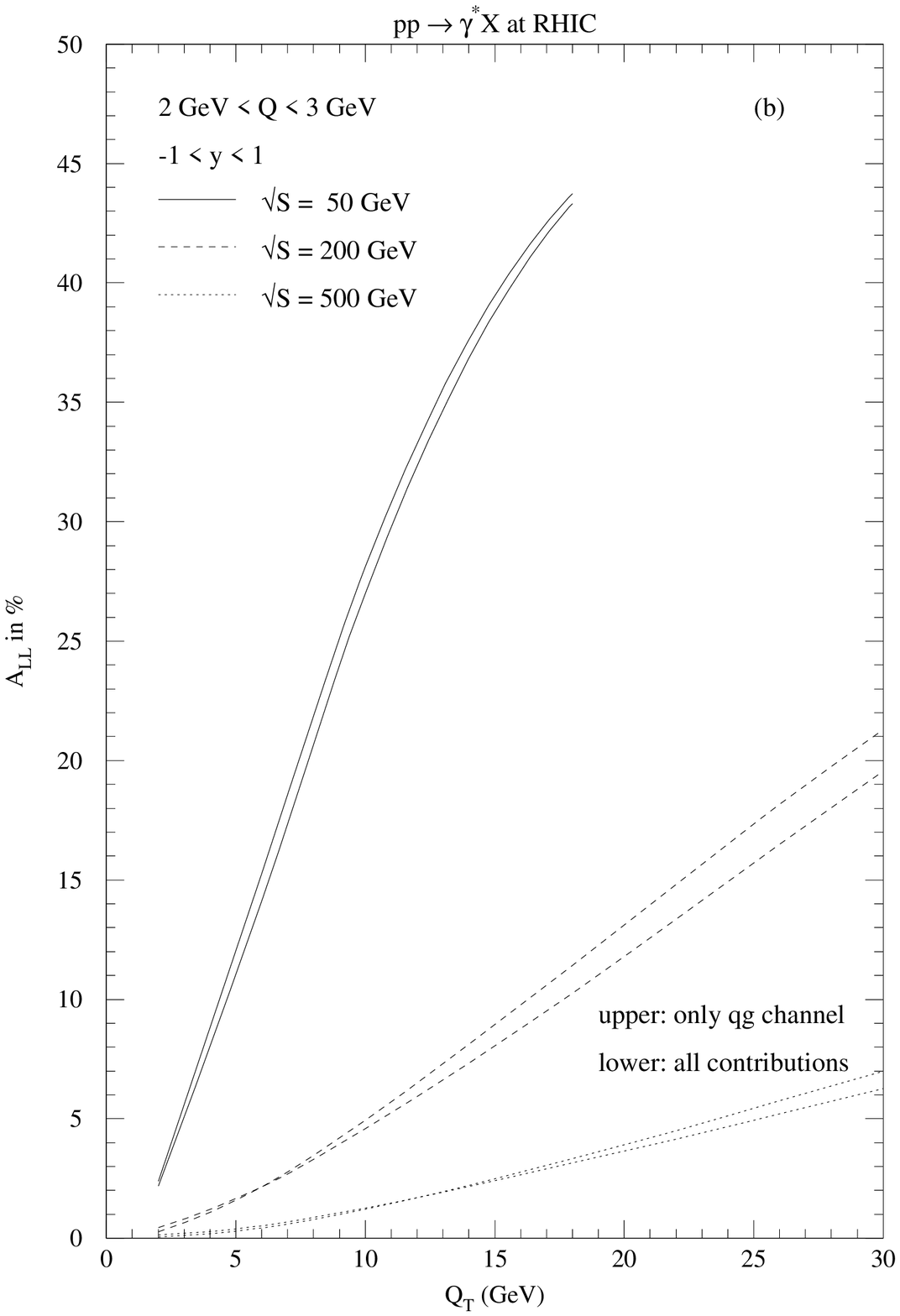,bbllx=55pt,bblly=100pt,bburx=495pt,bbury=725pt,%
           height=18cm}
  \end{picture}}
 \end{center}
\caption{Computed longitudinal asymmetry $\protect A_{LL}$ as a 
function of $Q_T$ for 
for $p p \rightarrow \gamma^* +X$ at $\protect\sqrt{S} =$
50, 200, and 500 GeV, averaged over the rapidity interval -1.0 $< y <$ 1.0 
and the mass intervals (a) 5.0 $<Q<$ 6.0 GeV and (b) 2.0 $<Q<$ 3.0 GeV.  Shown 
are both the complete 
answer at leading-order and the contribution from the $qg$ subprocess.  
The GSA set of polarized parton densities is used.}
\label{fig10}
\end{figure}

\begin{figure}
 \begin{center}
  {\unitlength1cm
  \begin{picture}(12,18)
   \epsfig{file=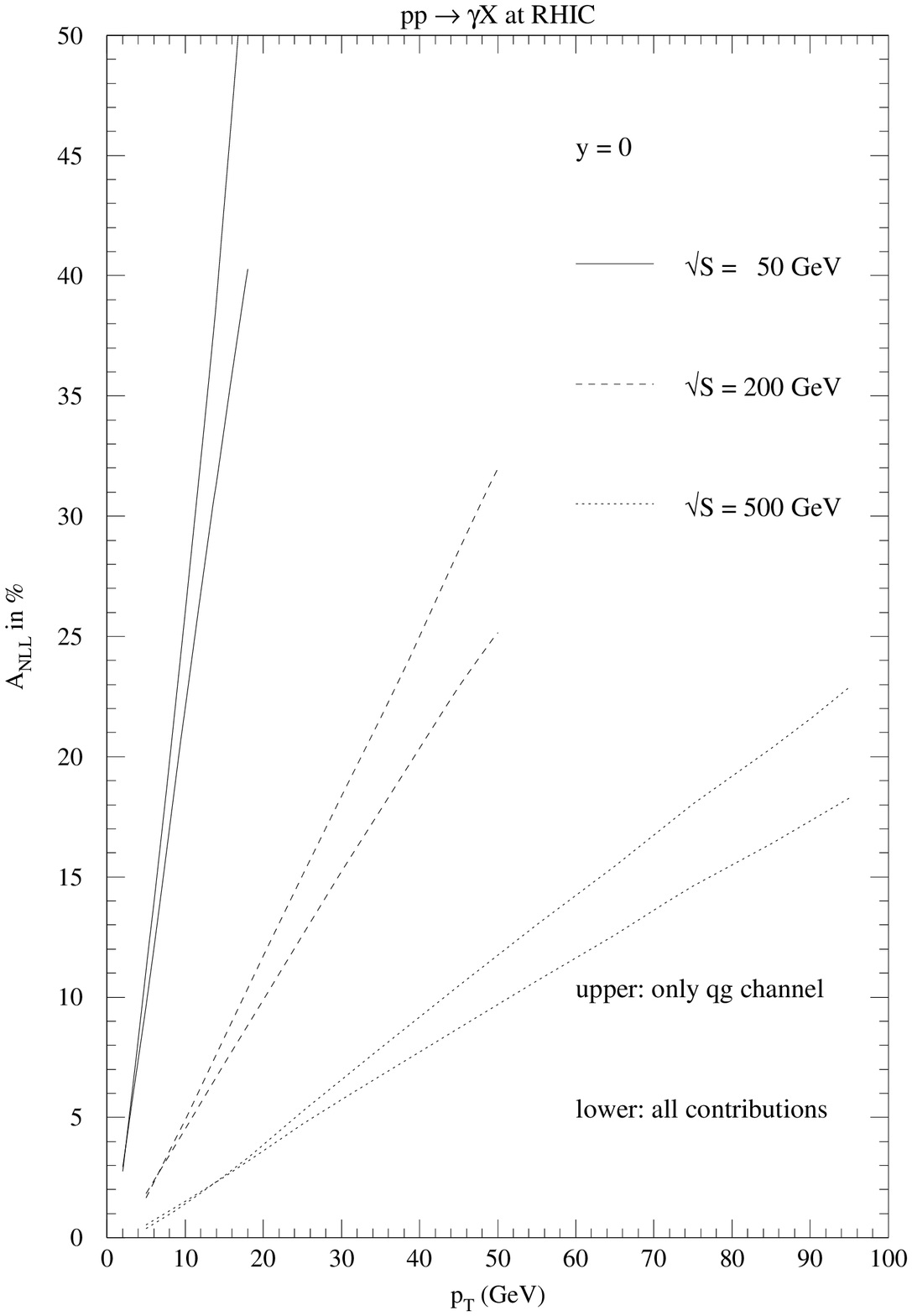,bbllx=55pt,bblly=100pt,bburx=495pt,bbury=725pt,%
           height=18cm}
  \end{picture}}
 \end{center}
\caption{Computed longitudinal asymmetry $\protect A_{LL}$ as a function 
of $p_T$ for real photon production $p p \rightarrow \gamma +X$ at  
$\protect\sqrt{S} =$ 50, 200, and 500 GeV and rapidity $ y = 0 $.   
Shown are both the complete 
answer at leading-order and the contribution from the $qg$ subprocess.  
The GSA set of polarized parton densities is used.}
\label{fig11}
\end{figure}



\begin{references}

\bibitem{ref:DY}
 S. Drell and T. M. Yan, Phys. Rev. Lett. {\bf 25}, 316 (1970); 
 Ann. Phys. (NY) {\bf 66}, 578 (1971).

\bibitem{ref:BQ} 
 E. L. Berger and J.-W. Qiu, Phys. Lett. {\bf B248}, 371 (1990) 
 and Phys. Rev. {\bf D44}, 2002 (1991).

\bibitem{ref:Baer}
 H.~Baer, J.~Ohnemus, and J.~F.~Owens, Phys. Rev. {\bf D42}, 61 (1990).

\bibitem{ref:Aurenche}
 P. Aurenche {\it{et al}}, Nucl. Phys. {\bf B399}, 34 (1993).

\bibitem{ref:GV}
 L.~E.~Gordon and W.~Vogelsang, Phys. Rev. {\bf D48}, 3136 (1993)
 and {\bf D50}, 1901 (1994); 
 M.~Gl\"uck, L.~E.~Gordon, E.~Reya, and W.~Vogelsang, Phys. Rev. Lett. 
 {\bf 73}, 388 (1994); 
 L. E. Gordon, Nucl. Phys. {\bf B501}, 175 (1997).

\bibitem{ref:BGKDY}
 E.~L.~Berger, L.~E.~Gordon, and M.~Klasen, Phys. Rev. {\bf D58}, 074012 (1998)
 and hep-ph/9906402.

\bibitem{ref:BGQ}
 E. L. Berger, X.-F. Guo, and J.-W. Qiu, Phys. Rev. Lett. {\bf 76}, 2234
(1996); Phys. Rev. {\bf D54}, 5470 (1996).
 E. L. Berger, X.-F. Guo, and J.-W. Qiu, in {\it'97 QCD and High Energy 
 Hadronic Interactions}, Proceedings of the XXXIInd Rencontres de Moriond, 
 Les Arcs, France, March, 1997, edited by J. Tran Thanh Van (Editions 
 Frontieres, Paris, 1997) pp 267 - 274.  

\bibitem{ref:Reno}
P. B. Arnold and M. H. Reno, Nucl. Phys. {\bf B319}, 37 (1989), 
and erratum Nucl. Phys. {\bf B330}, 284 (1990);
R. J. Gonsalves, J. Pawlowski, and C.-F. Wai, Phys. Rev. {\bf D40}, 2245 (1989).

\bibitem{ref:CSS}
J. Collins, D. Soper, and G. Sterman, Nucl. Phys. {\bf B250}, 199 (1985);
J. Collins and D. Soper, Nucl. Phys. {\bf B193}, 381 (1981),
                         Nucl. Phys. {\bf B197}, 446 (1982), and erratum
                         Nucl. Phys. {\bf B213}, 545 (1983).

\bibitem{ref:DWS}
C. Davies, B. Webber, and W. J. Stirling, Nucl. Phys. {\bf B256}, 413 (1985);
C. Davies and W. J. Stirling, Nucl. Phys. {\bf B244}, 337 (1984).

\bibitem{ref:AEGM}
G. Altarelli, R. K. Ellis, M. Greco, and G. Martinelli, Nucl. Phys. {\bf B246},
 12 (1984).  For a recent treatment and list of references, see R.~K.~Ellis,
 D.~A.~Ross, and S.~Veseli, Nucl. Phys. {\bf B503}, 309 (1997).

\bibitem{ref:AK}
P. B. Arnold and R. Kauffman, Nucl. Phys. {\bf B349}, 381 (1991).

\bibitem{ref:LY}
G. A. Ladinsky and C. P. Yuan, Phys. Rev. {\bf D50}, 4239 (1994).

\bibitem{ref:BerSop}
 E. L. Berger and D. E. Soper, Nucl. Phys. {\bf B247}, 29 (1984).  

\bibitem{ref:CCFG} S. Chang, C. Coriano, R. D. Field, and L. E. Gordon, 
Nucl. Phys. {\bf B512} 393, (1998); 
 S. Chang, C. Coriano, and R. D. Field, Nucl. Phys. {\bf B528} 285, (1998).

\bibitem{ref:Frixvogel} S. Frixione and W. Vogelsang, hep-ph/9908387.

\bibitem{ref:MRST} 
 A.~D.~Martin, R.~G.~Roberts, W.~J.~Stirling, and R.~S.~Thorne, 
 Eur.~Phys.~J. {\bf C4}, 463 (1998). 

\bibitem{GS} T.~Gehrmann and W.~J.~Stirling, Phys. Rev. {\bf D53}, 6100 (1996). 

\bibitem{ref:BergerGordon}
 E.~L.~Berger and L.~E.~Gordon, Phys. Rev. {\bf D58}, 114024 (1998).  

\end{references}
\end{document}